\documentclass[sigconf]{acmart}

\usepackage{graphicx} 
\usepackage{svg}
\usepackage{xcolor}
\usepackage{hyperref}
\usepackage{booktabs} 
\usepackage{algorithm}
\usepackage{algpseudocode}
\algrenewcommand{\algorithmiccomment}[1]{\hfill\emph{> #1}}
\usepackage{minted}
\newlength{\mintednumbersep}
\AtBeginDocument{%
  \sbox0{\tiny00}%
  \setlength\mintednumbersep{\leftskip}%
  \addtolength\mintednumbersep{-\wd0}%
}

\usepackage{tikz}

\usetikzlibrary{angles,quotes}
\usetikzlibrary{tikzmark}
\usetikzlibrary{arrows.meta}
\usepackage{siunitx}
\usepackage{todonotes}
\usepackage{forest}
\usepackage{balance}

\usepackage[a-1b]{pdfx}

\definecolor{dkgreen}{RGB}{0,64,0}
\definecolor{ltgray}{RGB}{245,245,245}
\definecolor{mauve}{RGB}{139,0,139}

\usepackage{caption}
\usepackage{subcaption}
\usepackage{enumerate}

\usepackage{cleveref}

\usepackage{multirow}
\usepackage{multicol}
\usepackage{array}
\usepackage{mwe}
\usepackage{keycommand}
\usepackage{fontawesome5}
\usepackage{tabularx}
\usepackage{rotating}
\selectcolormodel{natural} 
\usepackage{ninecolors} 
\selectcolormodel{rgb} 
\usepackage{tabularray}
\UseTblrLibrary{booktabs}

\makeatletter
\newcommand{\thickhline}{%
    \noalign {\ifnum 0=`}\fi \hrule height 1pt
    \futurelet \reserved@a \@xhline
}
\newcolumntype{"}{@{\hskip\tabcolsep\vrule width 1pt\hskip\tabcolsep}}
\makeatother

\definecolor{ceruleanblue}{rgb}{0.16, 0.32, 0.75}
\definecolor{lightblue}{rgb}{0.85, 0.92, 0.99} 

\definecolor{brightmaroon}{rgb}{0.76, 0.13, 0.28}
\definecolor{lightred}{rgb}{0.96, 0.86, 0.86} 

\definecolor{britishracinggreen}{rgb}{0.0, 0.26, 0.15}
\definecolor{lightgreen}{rgb}{0.78, 0.87, 0.78} 

\usepackage{listings}
  \lstdefinestyle{tree}{
    basicstyle=\ttfamily\color{black},
    moredelim=**[is][\color{brightmaroon}]{@}{@},
    moredelim=**[is][\color{ceruleanblue}]{&}{&},
    moredelim=**[is][\color{britishracinggreen}]{!}{!},    
    literate=
    {├}{{\smash{\raisebox{-1ex}{\rule{1pt}{\baselineskip}}}\raisebox{0.5ex}{\rule{1ex}{1pt}}}}1 
    {─}{{\raisebox{0.5ex}{\rule{1.5ex}{1pt}}}}1 
    {└}{{\smash{\raisebox{0.5ex}{\rule{1pt}{\dimexpr\baselineskip-1.5ex}}}\raisebox{0.5ex}{\rule{1ex}{1pt}}}}1 
  }

  \lstdefinestyle{plaincodetextstyle}{
    basicstyle=\fontsize{8}{8}\ttfamily\color{black}, 
    numbers=left,  
    numberstyle=\tiny\color{gray},   
    stepnumber=1,  
    backgroundcolor=\color{ltgray}, 
    showstringspaces=false,  
    tabsize=4,       
    breaklines=true, 
    breakatwhitespace=true, 
    escapeinside=||,
    framexleftmargin=14pt, 
    xleftmargin=10pt, 
    aboveskip=4pt, 
}

\newcommand{\eg}{{\textit{e.g., }}}
\newcommand{\ie}{{\textit{i.e., }}}

\newcommand{\rajaperfabbrev}{RAJAPerf}
\newcommand{\rajaperf}{RAJA Performance Suite}

\newcommand{\altimes}{Apps\_LTIMES}
\newcommand{\kltimes}{Kripke\_LTIMES}

\newcommand{\kmeans}{k-means}
\newcommand{\agg}{agglomerative}
\newcommand{\topdown}{top-down}
\newcommand{\ncu}{ncu}

\usepackage{tikz}
\usepackage{graphicx} 
\usepackage{subcaption}

\usepackage{cuted}
\usepackage{caption}

\usetikzlibrary{positioning}

\newif\ifdraft{}
\drafttrue{}

\AtBeginDocument{%
  }

\setcopyright{none}

\usepackage{enumitem}
\setlist[itemize]{leftmargin=5mm}

\begin{document}

\everypar{\looseness=-1}

\newcommand{\LLNL}{{LLNL}}
\newcommand{\caliper}{Caliper}
\newcommand{\EDA}{EDA}

\newcommand{\ramble}{Ramble}
\newcommand{\spack}{Spack}

\newcommand{\saxpy}{\tt saxpy}

\title[]{On Similarity of Computational Kernels in our Codes and Proxies}

\author{Michael McKinsey}
\email{mckinsey1@llnl.gov}
\affiliation{%
  \institution{Lawrence Livermore National Lab}
  \city{Livermore}
  \state{California}
  \country{USA}
}

\author{Stephanie Brink}
\email{brink2@llnl.gov}
\affiliation{%
  \institution{Lawrence Livermore National Lab}
  \city{Livermore}
  \state{California}
  \country{USA}
}

\author{Olga Pearce}
\email{pearce8@llnl.gov}
\affiliation{%
  \institution{Lawrence Livermore National Lab}
  \city{Livermore}
  \state{California}
  \country{USA}
}
\affiliation{%
  \institution{Texas A\&M University}
  \city{College Station}
  \state{Texas}
  \country{USA}
}


\renewcommand\_{\textunderscore\allowbreak}

\begin{abstract}

As high-performance computing (HPC) systems rapidly evolve, with increasing on-node parallelism and widespread use of accelerators, understanding how the code maps to hardware is essential for reaching optimal performance. 
Benchmarks are commonly used for early assessment of emerging architectures (as well as for informing the design of future hardware), but it is often unknown how well the benchmarks represent the performance characteristics of simulation codes.
Existing methods for evaluating how well our benchmarks represent our HPC codes are manual, labor intensive, and challenging to scale to many benchmarks. 
In this paper, we propose performance similarity metrics based on how the code uses the compute hardware. 
We define and characterize two broad categories of kernels that exhibit similar performance characteristics.
We evaluate the pairwise similarity metrics on kernels in the Kripke proxy application and the \rajaperf, 
using both a CPU-only system and a GPU-accelerated system. 
We validate that our similarity metrics correctly match a kernel in the Kripke proxy app to a kernel in the \rajaperf.  
Our proposed similarity metrics enable assessment of the similarity of computational kernels in our codes 
and the proxy applications we use to represent the codes.


\end{abstract}





\maketitle

\section{Introduction}

High-performance computing (HPC) architectures and accelerator technologies are changing quickly to continue to add parallelism on the node, driven by
the needs of scientific computing and the AI race.
Taking full advantage of the performance of these architectures requires an in-depth understanding of the mapping of the code to the hardware.
The HPC community has come to rely on benchmarks for an initial quick 
assessment of the performance of the new hardware. 
However, are these benchmarks representative of the compute we have in our simulations?

State-of-the-art approaches to evaluating benchmark representation 
are manual and tedious. A report by the ECP Proxy App Project~\cite{richards:ecp2021} highlights the large amount of 
manual effort currently required to evaluate a single simple proxy application. Assessing similarities between benchmarks and codes
remains an active area of research with approaches 
to compare spatial temporal locality of memory 
accesses~\cite{f0f3240e4a9510cc07515899834b8e7b122f1a38}
and matching signatures~\cite{matching-signatures}.

In this paper, we propose performance similarity metrics to determine 
whether two kernels are similar in how they utilize a given hardware.
We study the similarity of kernels in the \rajaperf, 
and the Kripke proxy application.
We define and characterize two broad categories of kernels, \emph{memory bound} and \emph{compute bound}.
We provide analysis on a CPU system with Intel's Sapphire Rapids CPUs, 
and a system accelerated with NVIDIA's H100 GPUs. The main contributions of our work are as follows:

\begin{itemize}
\item Identify hardware metrics which can serve as the basis for 
    understanding code performance on CPUs and GPUs (Section~\ref{sec:hardware_metrics}).
\item Propose code performance similarity metrics 
    (Section~\ref{sec:similarity_metrics}).
\item Validate the proposed performance similarity
    metrics on Kripke and the \rajaperf \
    on CPUs and GPUs (Sections~\ref{sec:agglomerative_kmeans},~\ref{sec:kripke_vs_rajaperf}).
\item Propose a methodology to select the number of broad categories of kernels in Section~\ref{sec:selection}.
\end{itemize}
\section{Background}

In this section, we introduce the benchmarks and system configurations used in this work. 
We also detail the hardware performance metrics for the CPU and GPU architectures.

\subsection{Benchmarks}
\label{sec:benchmarks}

In this paper, we study the Kripke proxy application and the \rajaperf \ (\rajaperfabbrev). Kripke~\cite{kripke,kunen2015kripke} is a deterministic transport proxy application designed to study the performance characteristics of data layouts, programming models, and sweep algorithms~\cite{kripke,kunen2015kripke}.
Kripke uses MPI for network communication and a performance portability programming model known as RAJA~\cite{beckingsaleRAJAPortablePerformance2019}
for on-node parallelism.  
Kripke can run on CPU systems, and systems accelerated with NVIDIA or AMD GPUs.
Kripke is highly parameterizable to enable performance exploration.

\begin{figure}[thb]
    \includegraphics[width=\columnwidth]{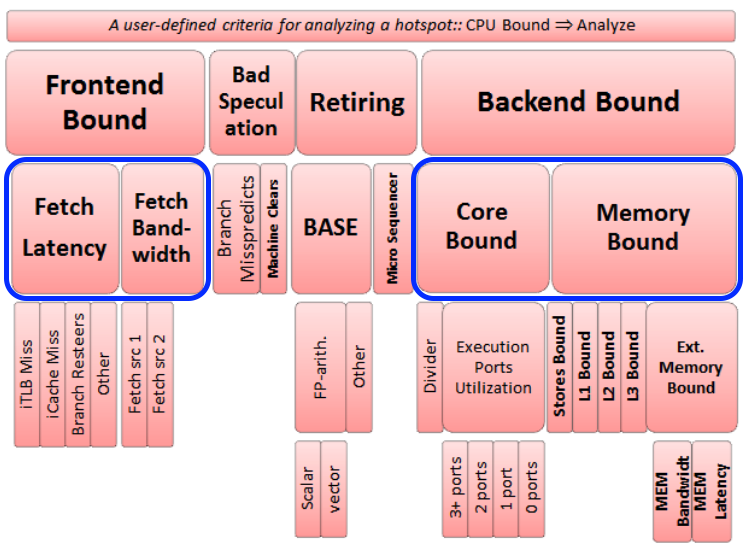}
    \vspace{-0.7cm}
    \caption{Top-down hierarchy on Intel CPUs. In this study, we use the four subcategories highlighted in blue.}
    \label{fig:topdown}
        \vspace{-0.25cm}
\end{figure}

The \rajaperf~\cite{llnlRAJAPerf2024,10.1145/3754598.3754668,10.1145/rajaperf-2026,10.1007/978-3-031-97635-3_39} 
is a curated collection of computational kernels derived from standardized benchmarks and HPC applications. These kernels serve as fundamental building blocks of HPC simulations, enabling targeted performance evaluation and optimization.
The kernels in \rajaperfabbrev \ are implemented using RAJA, and
can therefore run on the same platforms as Kripke.
\rajaperfabbrev \ offers command-line options to adjust parameters such as \textit{tuning} (\eg block size) and \textit{problem size}, enabling flexible performance exploration. \rajaperfabbrev \ provides basic performance metrics, such as bytes read or written and FLOP counts for each kernel.  

Both \rajaperfabbrev \ and Kripke are annotated with Caliper~\cite{boehme:2016:caliper,caliper},
enabling both timing and hardware counter collection.
Kripke and \rajaperfabbrev \ both contain the LTIMES kernel,
which we use to evaluate our proposed performance similarity metrics.
Our goal when comparing Kripke and \rajaperfabbrev \ is to correctly match the LTIMES kernels.

We run 46 
kernels in \rajaperfabbrev \ for this study. We provide a detailed description of the kernels and kernel 
categories in previous work~\cite{llnlRAJAPerf,10.1145/3754598.3754668}.
The number of kernels per category that we include in this analysis and their category description is as follows:
\begin{itemize}
\item Algorithms (2): Kernels that focus on specific parallel constructs.
\item Applications (13): Kernels derived from important application
operations in various LLNL multiphysics applications.
\item Basic Patterns (10): Kernels that are small and simple yet
often present optimization challenges for compilers.
\item LCALS (9): Kernels originating in the Livermore Loops
suite~\cite{llnl_loops86}.
\item Polybench (8): A subset of kernels from the Polybench Suite
used to study polyhedral optimization in compilers~\cite{polybench}.
\item Stream (4): Streaming kernels found in the McCalpin STREAM benchmark~\cite{McCalpin2007}.
\end{itemize}

\subsection{Experimental Setup}
\label{sec:setup}

We run experiments on the Dane (CPU) and Matrix (GPU) systems at LLNL. 
Dane nodes have two sockets with Intel Sapphire Rapids CPUs (Xeon Platinum 8480+) with 56 cores each and 256 GB of DDR5 node memory.
Matrix nodes have four NVIDIA H100 GPUs with 80 GB of memory per GPU and 512 GB of CPU memory. 

CPU runs are three trials on an exclusive single node using all 112 cores available in the node. 
GPU runs are a single trial on an exclusive single node profiled from a single H100.
For each trial, kernels are repeated multiple times to mitigate noise.
We leverage Benchpark~\cite{hpctests2023-benchpark,benchpark,10.1145/3736731.3746150} to ensure reproducible results, including building the benchmarks and running all of the experiments necessary to generate results for this study.
We use Thicket~\cite{thicket,brink2023thicket} to compose the multi-dimensional Caliper performance data, 
perform exploratory data analysis (EDA), and clustering. 


\begin{figure}[htb]
    \centering
    \includegraphics[width=\columnwidth]{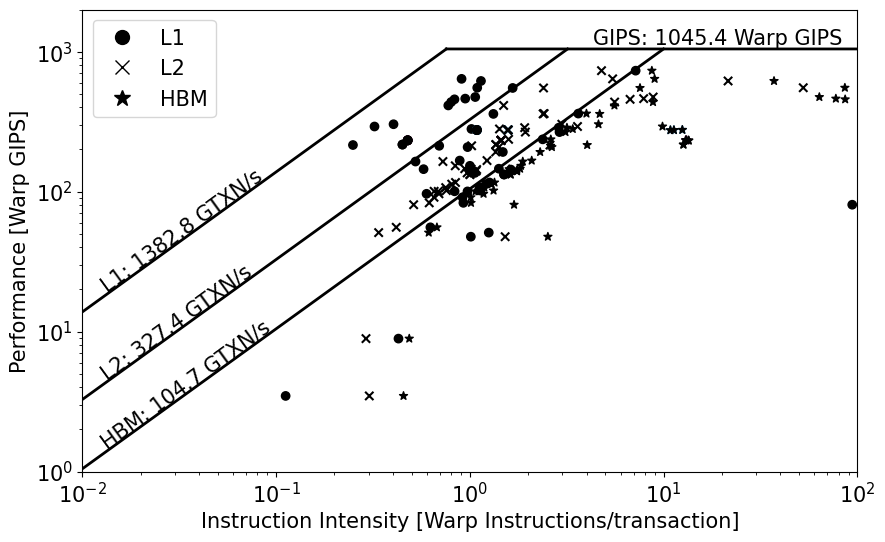}
    \vspace{-0.7cm}
    \caption{GPU roofline analysis of \rajaperf \ kernels on NVIDIA H100 GPU.}
    \vspace{-0.25cm}
\label{fig:h100-roofline}
\end{figure}



\begin{figure*}[htb]
    \centering
    \begin{subfigure}[t]{\columnwidth}
        \centering
        \includegraphics[width=\columnwidth]{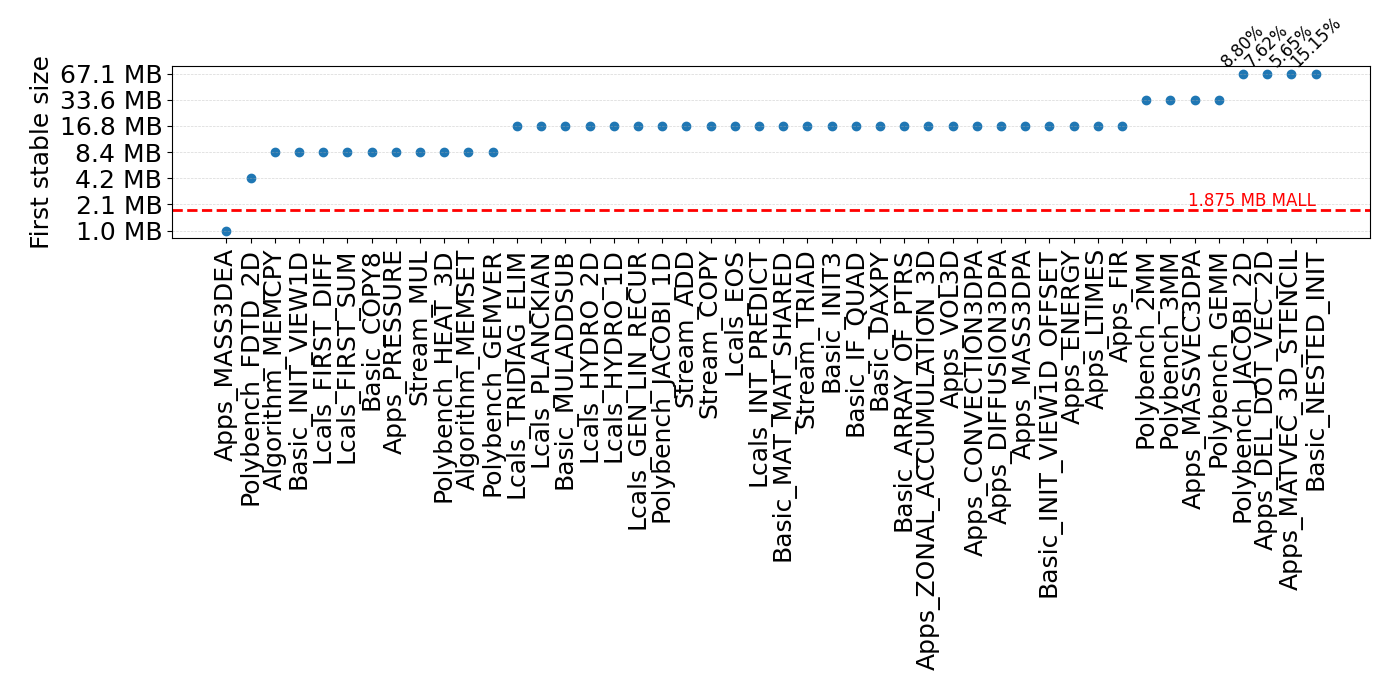}
            \vspace{-0.75cm}
        \caption{Min. memory/core for stable \topdown\ metrics (Sapphire Rapids)}
        \label{fig:stable-size}
    \end{subfigure}
    \hfill
    \begin{subfigure}[t]{\columnwidth}
        \centering
        \includegraphics[width=\columnwidth]{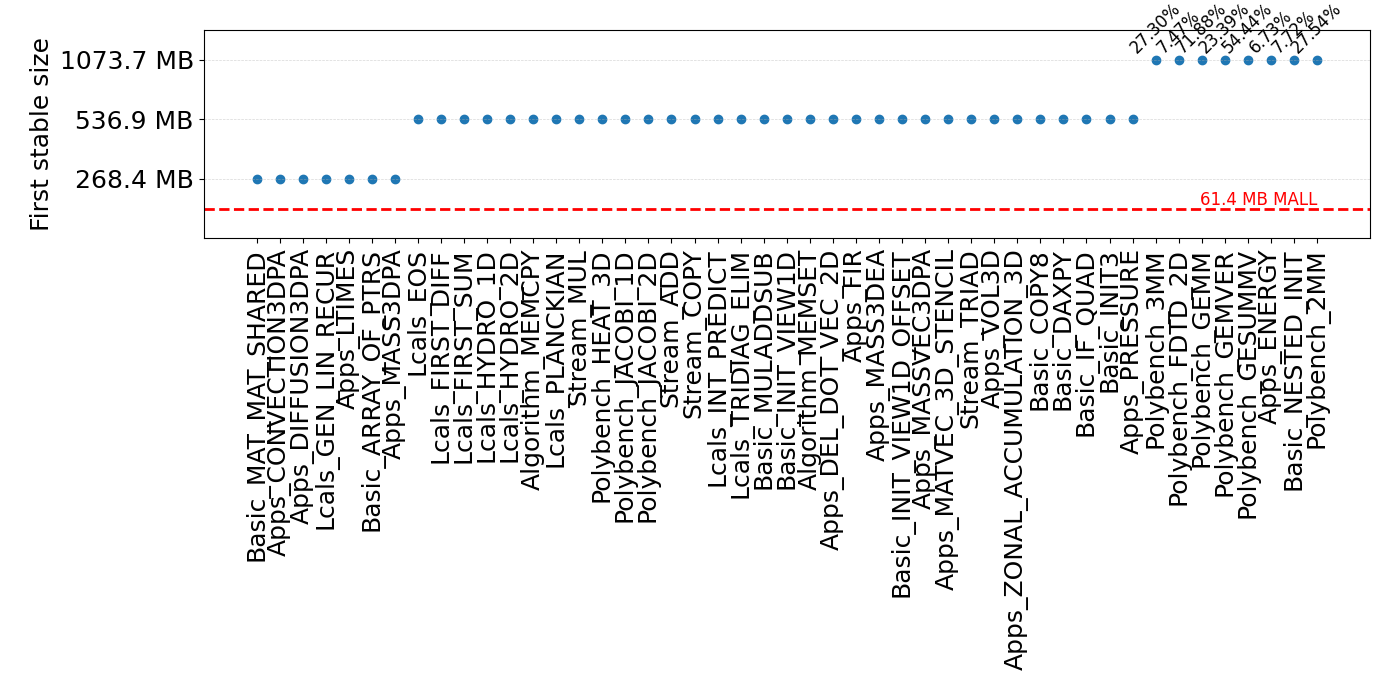}
            \vspace{-0.75cm}
        \caption{Min. memory/device for stable \ncu\ metrics (H100)}
        \label{fig:stable-size-gpu}
    \end{subfigure}
    \vspace{-0.25cm}
    \caption{\label{fig:stable-cpu-gpu}Minimum memory size for stable performance metrics (\eg <5\% difference with larger problem sizes) for RAJA Performance Suite kernels.
    We annotate kernels that are never stable (\eg they have at least one metric with >5\% difference) with its largest percent difference across the metrics.}
        \vspace{-0.25cm}
\end{figure*}

\subsection{Hardware Metrics}
\label{sec:hardware_metrics}

We evaluate two high-level methodologies for characterizing code performance,
Intel's \topdown \ analysis on CPUs~\cite{yasin2014top} and a subtle variation of roofline analysis on GPUs~\cite{gpu_rooflines}.
The \topdown \ methodology identifies performance bottlenecks in the CPU pipeline, specifically in four main categories shown in Figure~\ref{fig:topdown}, frontend bound, bad speculation, retiring, and backend bound.
Drilling down into any single category helps uncover performance bottlenecks in the application. In this study, we use the following four subcategories:
\begin{enumerate}
    \item Core bound - stalls representing either short execution starvation periods, or sub-optimal execution ports utilization
    \item Memory bound - stalls related to the memory subsystem
    \item Fetch latency - corresponds to instruction cache misses
    \item Fetch bandwidth - inefficiency in the instruction decoders
\end{enumerate}
Core bound and memory bound are subcategories of the top-level backend bound category, whereas fetch latency and fetch bandwidth are subcategories of the top-level frontend bound category.
We do not use the bad speculation category, as none of the kernels in the \rajaperf\ exhibit irregular memory access patterns that would attribute to this category.
Additionally, we omit the retiring category because it is the remainder of the other topp-level categories and therefore does not add information.

Roofline analysis on GPUs characterizes kernel performance in GigaFLOPs (GFLOPS/s) as a function of the kernel's arithmetic intensity~\cite{gpu_rooflines}, which is a ratio of floating-point operations performed to data movement (see Figure~\ref{fig:h100-roofline}).
We use Caliper to collect GPU runtime, and NVIDIA's Nsight Compute (\ncu)~\cite{ncu} to capture hardware metrics which include the number of total warp instructions, L1 transactions, L2 transactions, HBM transactions, as demonstrated in our prior work~\cite{10.1145/3754598.3754668}. 
We substitute the arithmetic intensity metrics with transaction rate metrics. 
The three arithmetic intensity metrics are derived from warp instructions and therefore are highly correlated with the IPS metric.
The transaction rate metrics use counters other than warp instructions,  thus are not redundant to the IPS metric and provide additional information.
We use the following four derived metrics:
\begin{enumerate}
    \item L1 transaction rate = $L1_{\text{Transactions,total}} / Time_{\text{GPU}}$
    \item L2 transaction rate = $L2_{\text{Transactions,total}} / Time_{\text{GPU}}$
    \item HBM transaction rate = $HBM_{\text{Transactions,total}} / Time_{\text{GPU}}$
    \item Performance (IPS) = $\text{Instructions}_{\text{warp}} / Time_{\text{GPU}}$
\end{enumerate}
We compute the transaction rate instead of using total transactions, as runtime directly influences the number of transactions.

Different workload problem sizes may result in different values of the hardware metrics.
For example, a problem size that fits into cache will have fewer cache line evictions than
a problem size that does not fit into cache, and therefore lower cache line misses.
Cache line misses will influence the memory bound metric on CPU and the cache (L1/L2/HBM) transaction metrics on the GPU.
We analyze these hardware metrics differences to select a sufficiently large problem sizes.
Figure~\ref{fig:stable-cpu-gpu} provides a summary of the variation in hardware metrics observed in 
the \topdown \ and \ncu \ datasets for all \rajaperfabbrev \ kernels.
An increase memory usage on the y-axis directly maps to a larger problem size for each kernel, 
with the difference in problem size depending on the kernel.
We observe that values of the \topdown \ and \ncu \ metrics are highly dependent on the problem size.
Figure~\ref{fig:stable-size} shows the minimum memory size for stable performance metrics 
for all kernels in \rajaperfabbrev. The dotted line at 1.875 MB indicates the size of the 
Sapphire Rapids memory-attached last-level cache (MALL). When running at 2.1 MB---a size that is 
greater than the CPU MALL---only one kernel out of 46 has less than 5\% performance variation 
(stable performance) when comparing performance data to a larger size.
We observe that a problem size of at least 33.6 MB is required for 42 
out of 46 kernels to have stable \topdown \ values.

Similar to the CPU, only a handful of kernels (seven out of 46) are stable on the GPU (Figure~\ref{fig:stable-cpu-gpu}), even when running at 268.4 MB. 
Note that the H100 MALL size is 61.4 MB (dotted lne). 
At 1 GB, we see that 38 out of 46 kernels stabilize, but there is still up to 71.8\% difference between the last two problem sizes.
We focus our analysis on the largest problem sizes for each platform, which are the most stable.
\section{Performance Similarity Metrics}
\label{sec:similarity_metrics}

Our goal is to assess the similarity of kernels in terms of their performance on specific hardware. 
We need similarity metrics to quantify the similarity of (1) two kernels, and (2) a set of kernels.  
We represent each kernel by a list of hardware metrics we collect during the kernel's execution.
Each hardware metric is a single dimension in the $n$-dimensional hardware metric space, where $n$ is 
the number of hardware metrics we are using.
We define the {\bf distance} between kernels $p$ and $q$ as their $n$-dimensional Euclidean distance: 
\begin{equation}\label{eq:dist}
d(p, q) = \sqrt{(p_1 - q_1)^2 + (p_2 - q_2)^2 + \cdots + (p_n - q_n)^2}
\end{equation} 
where $p_i$ is the value of the $i$-th metric in the hardware metrics list for kernel $p$. 

We use clustering to group kernels into subsets (clusters) with similar performance characteristics.
Clustering automatically groups similar data points into clusters,
ensuring the items in each cluster are as similar as possible,
and that the clusters are distinct from each other.
Clustering methods use the list of hardware metrics per kernel as the input and the distance in Equation~\ref{eq:dist} to partition the kernels into compact, separable clusters. In this paper, we compare two clustering methods, \agg\ and \kmeans\ clustering. 

\subsection{Standardizing the Data}
\label{sec:standardizing}
To prevent certain metric values from dominating in the distance-based algorithms, we
standardize our dataset by removing the mean and scaling to unit variance for each metric.
Additionally, for metrics which span multiple orders of magnitude, 
we apply a natural log transformation before applying the standardization, 
to make the skewed data more normally distributed.
Note that similarity analysis in Section~\ref{sec:agglomerative_kmeans} and 
Section~\ref{sec:kripke_vs_rajaperf} is based on distances in the standardized dataset, 
not on the raw values of the original dataset.

\subsection{Agglomerative Clustering}
\label{sec:agglomerative_clustering}
We leverage the Scikit-learn implementation of hierarchical \agg \ 
clustering~\cite{scikitlearnAgglom}, using the Euclidean distance to compute the 
Ward linkage~\cite{ward_strat_agglom}. 
We visualize the clusters with a dendrogram, 
where the distance at which clusters are linked represents the Euclidean distance between cluster centroids.
We visualize the distribution of hardware metric values in each cluster with a barplot.

\subsection{K-means Clustering}
\label{sec:kmeans_clustering}
We use the Scikit-learn implementation~\cite{scikitlearnKmeans} of Lloyd's algorithm~\cite{kmeans}, with k-means++ initialization~\cite{10.5555/1283383.1283494}.  
Due to the number of dimensions in the datasets, we use PCA to visualize $n$-dimensional cluster centroids
in two dimensions.

\subsection{Evaluation}
\label{sec:eval}
We evaluate the resulting clusters by assessing the clusters' compactness, 
cluster separation, and the assignment of kernels to clusters as follows:

\subsubsection{Partitioning}
How many clusters the dataset should be partitioned into 
(discussion deferred to Section~\ref{sec:selection}), and
whether kernels are assigned to correct clusters
(Sections
~\ref{sec:cpu_partitioning},~\ref{sec:gpu_partitioning},~\ref{sec:cpu_gpu_partitioning}, 
and~\ref{sec:kripke_vs_rajaperf}).

\subsubsection{Compactness}
The average Euclidian distance of each kernel to 
the cluster centroid (Sections~\ref{sec:cpu_compactness},~\ref{sec:gpu_compactness},~\ref{sec:cpu_gpu_compactness}).

\subsubsection{Separation}
The average Euclidian distance between cluster centroids.
(Sections~\ref{sec:cpu_separation},~\ref{sec:gpu_separation},~\ref{sec:cpu_gpu_separation}).

\begin{figure}[htb]
    \centering
    \begin{subfigure}{\columnwidth}
        \centering
        \includegraphics[width=.925\columnwidth]{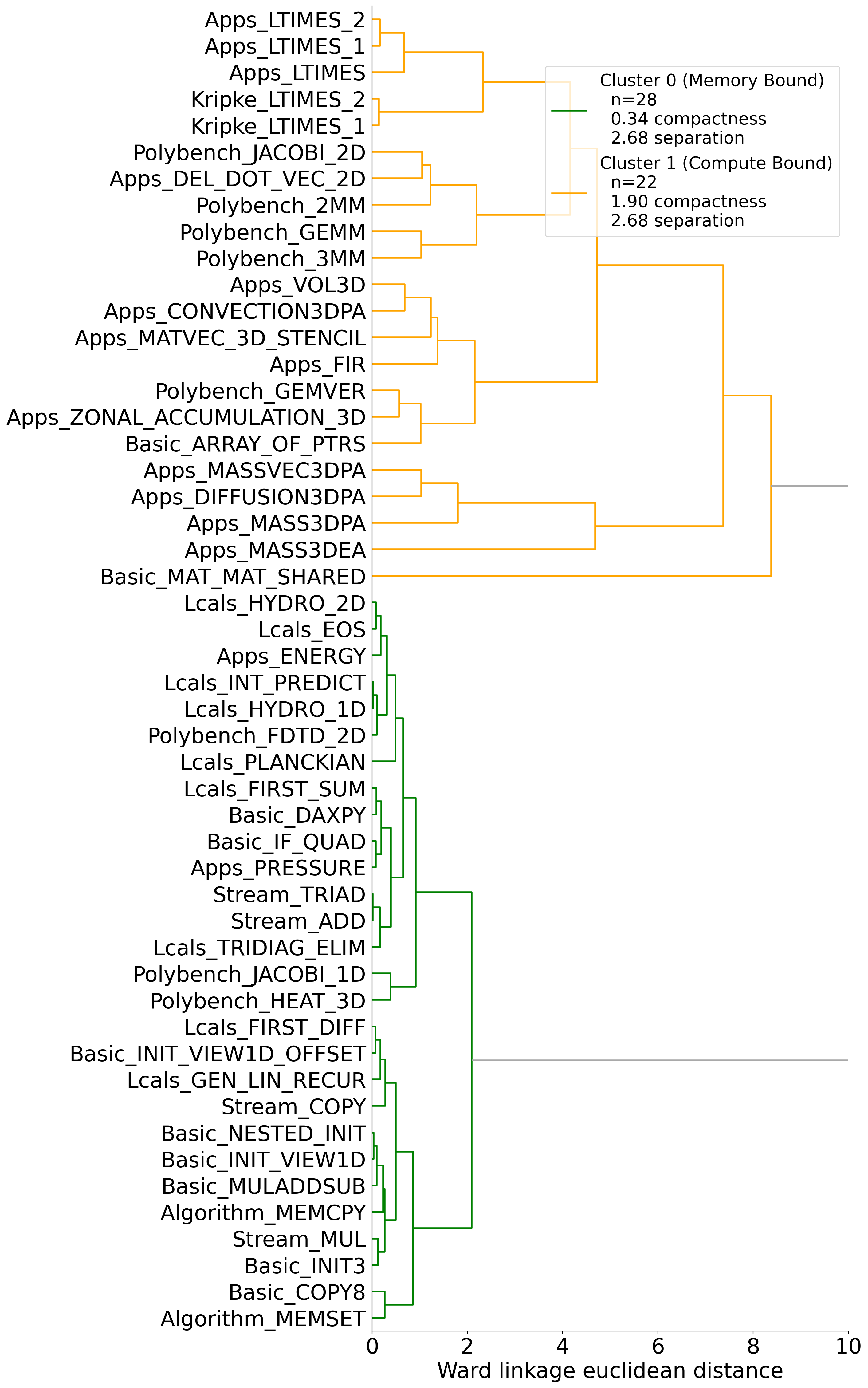}
        \vspace{-0.25cm}
        \caption{Dendrogram visualization of agglomerative clustering
        }     
        \label{fig:topdown_agg_dendrogram}
    \end{subfigure}
    \begin{subfigure}{\columnwidth}
        \centering
        \includegraphics[width=\columnwidth]{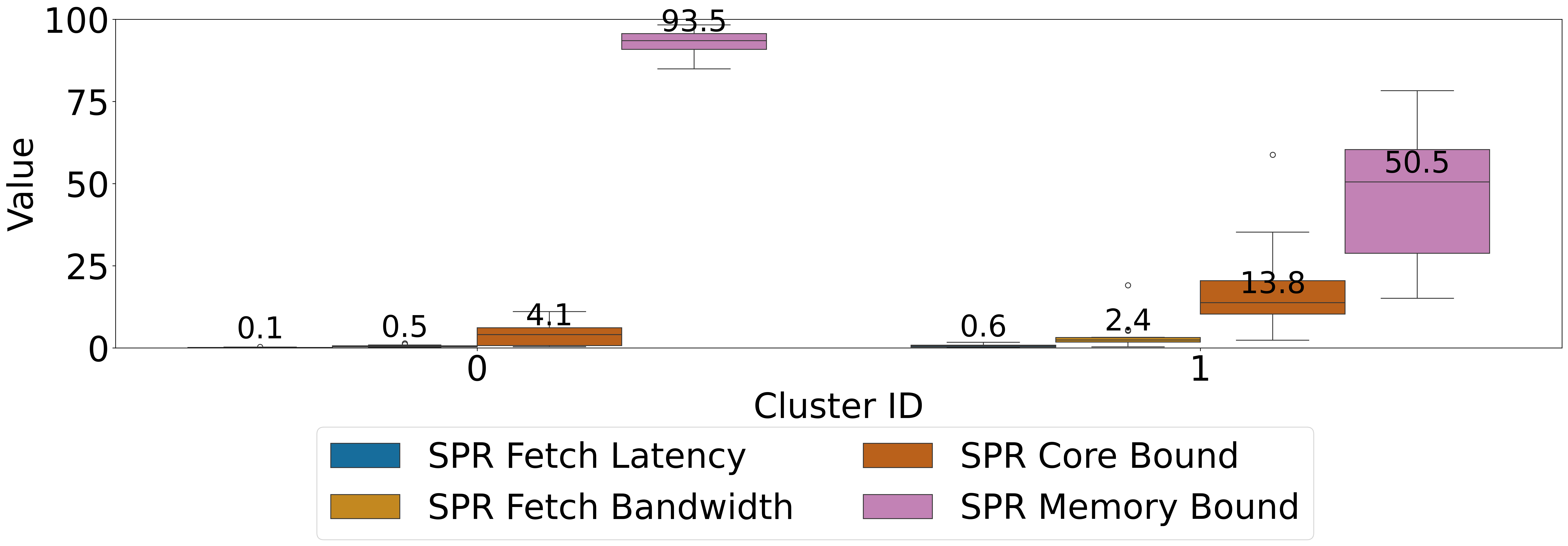}
        \caption{Distribution of metric values for each cluster}
        \label{fig:topdown_agg_barplot}
    \end{subfigure}
    \vspace{-0.75cm}
    \caption{Agglomerative clustering using \topdown \ metrics}
    \label{fig:topdown_agg}
    \vspace{-0.35cm}
\end{figure}
\begin{figure}[htb]
    \centering
    \begin{subfigure}{\columnwidth}
        \centering
        \includegraphics[width=\columnwidth]{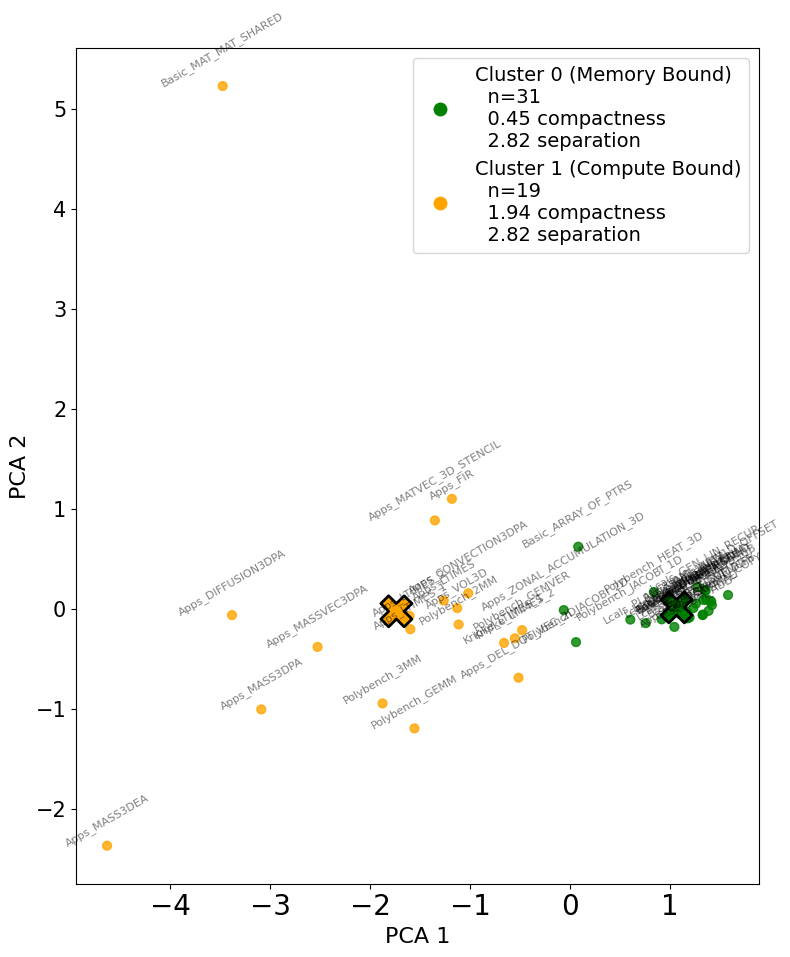}
        \caption{2D PCA visualization of $n$-dimensional k-means clustering}
        \label{fig:topdown_kmeans_scatter}
    \end{subfigure}
    \begin{subfigure}{\columnwidth}
        \centering
        \includegraphics[width=\columnwidth]{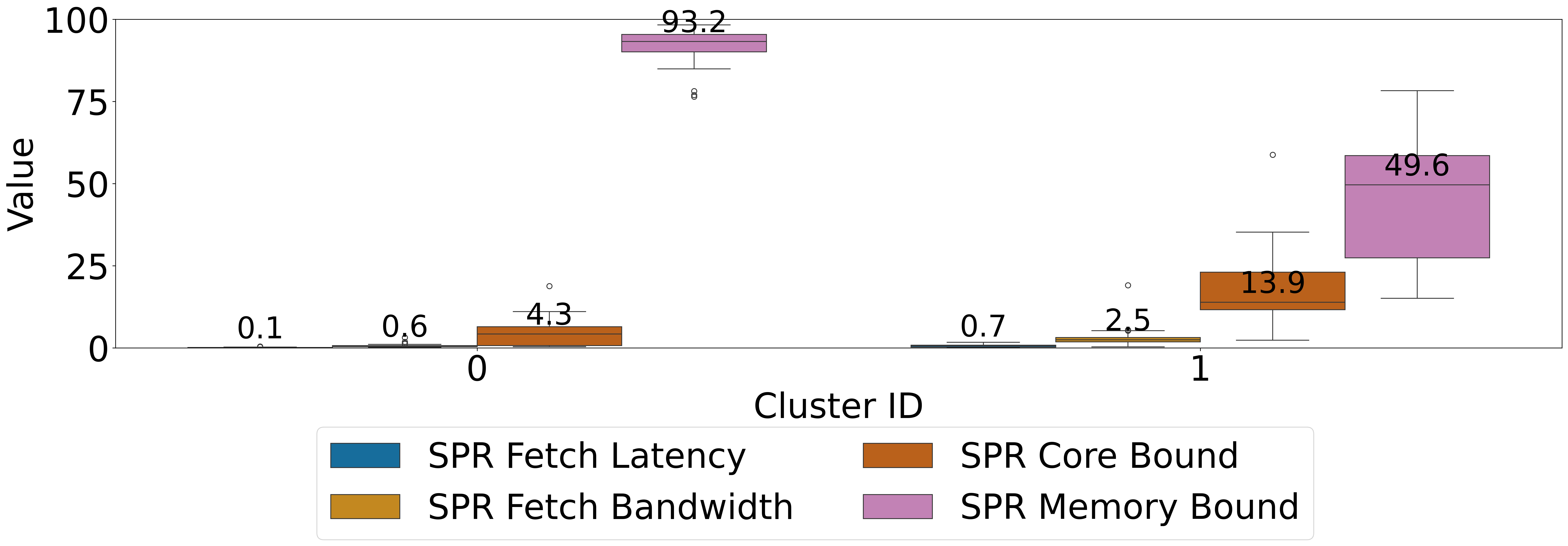}
        \caption{Distribution of metric values for each cluster}
        \label{fig:topdown_kmeans_barplot}
    \end{subfigure}
    \caption{K-means clustering using \topdown \ metrics}
    \label{fig:topdown_kmeans}
        \vspace{-0.25cm}
\end{figure}
\begin{figure}[htb]
    \centering
    \begin{subfigure}{\columnwidth}
        \centering
        \includegraphics[width=.925\columnwidth]{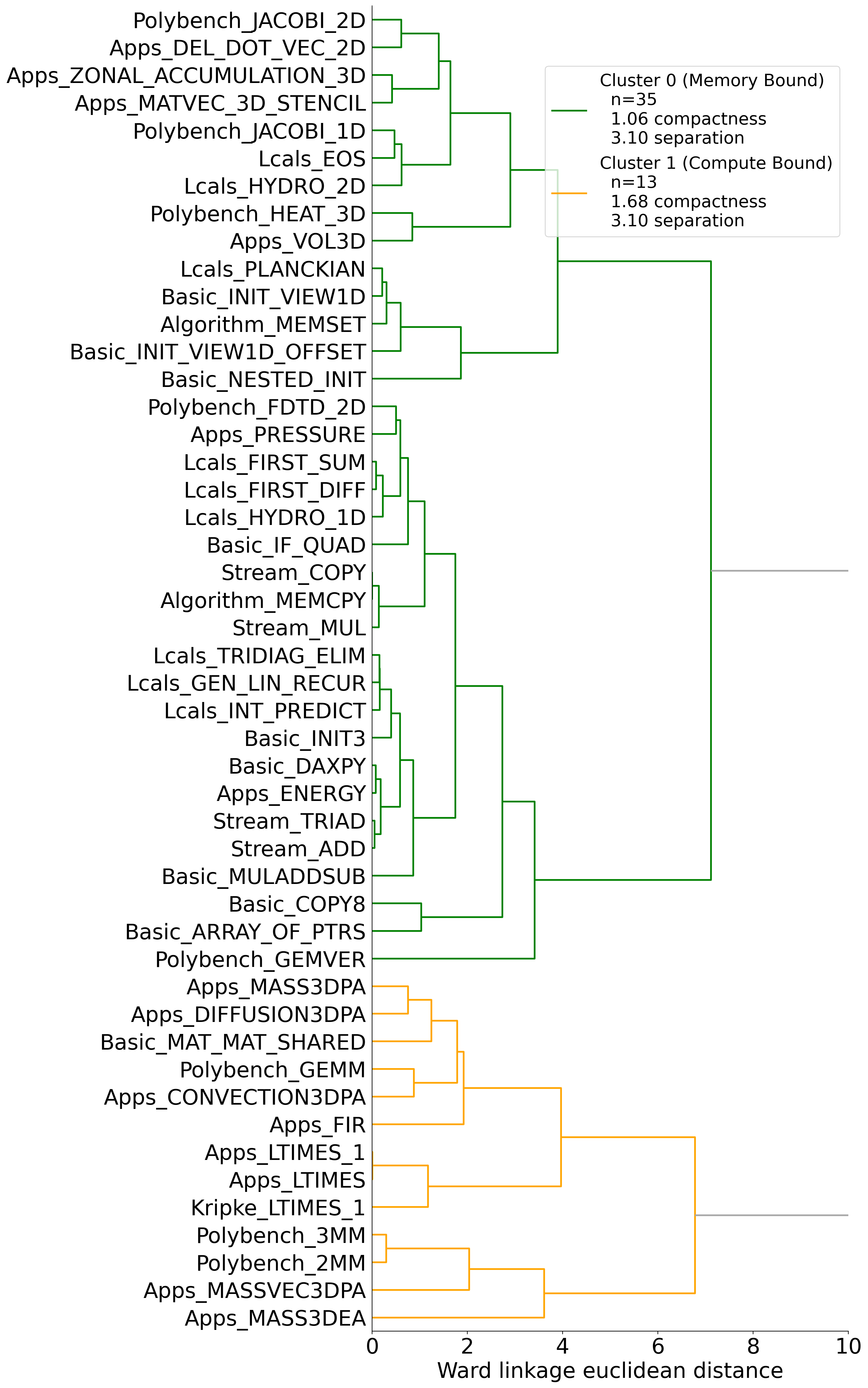}
        \caption{Dendrogram visualization of agglomerative clustering
        }
        \label{fig:ncu_agg_dendrogram}
    \end{subfigure}
    \begin{subfigure}{\columnwidth}
        \centering
        \includegraphics[width=\columnwidth]{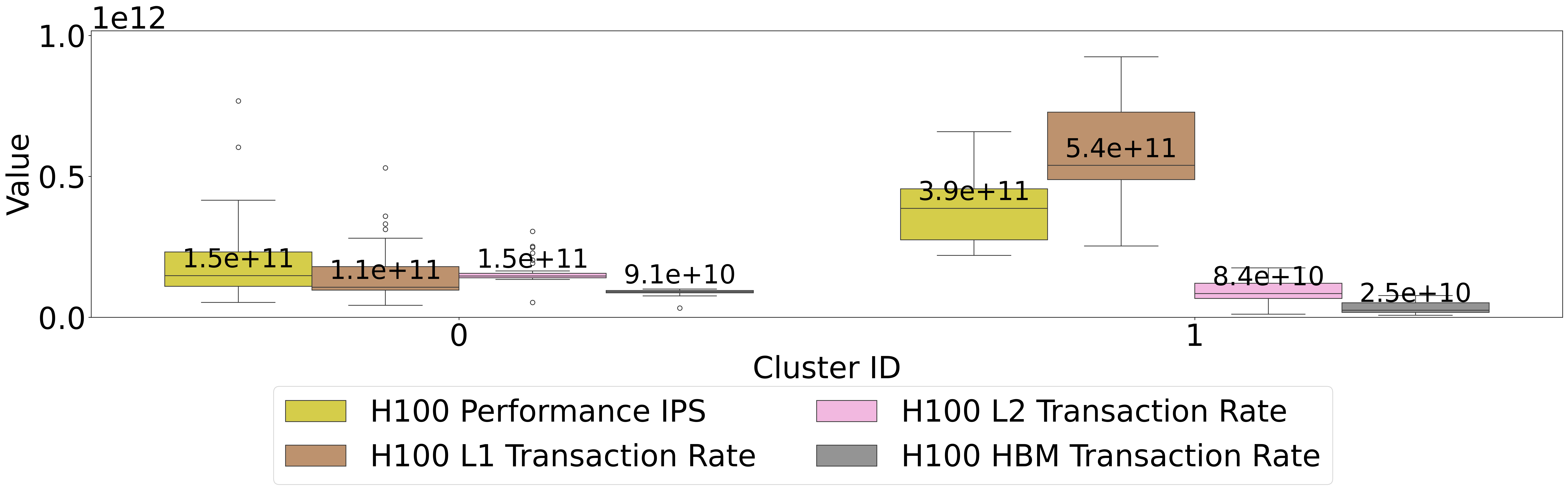}
        \caption{
        Distribution of metric values for each cluster}
        \label{fig:ncu_agg_barplot}
    \end{subfigure}
    \caption{Agglomerative clustering using \ncu \ metrics}
    \label{fig:ncu_agg}
\end{figure}
\begin{figure}[htb]
    \centering
    \begin{subfigure}{\columnwidth}
        \centering
        \includegraphics[width=\columnwidth]{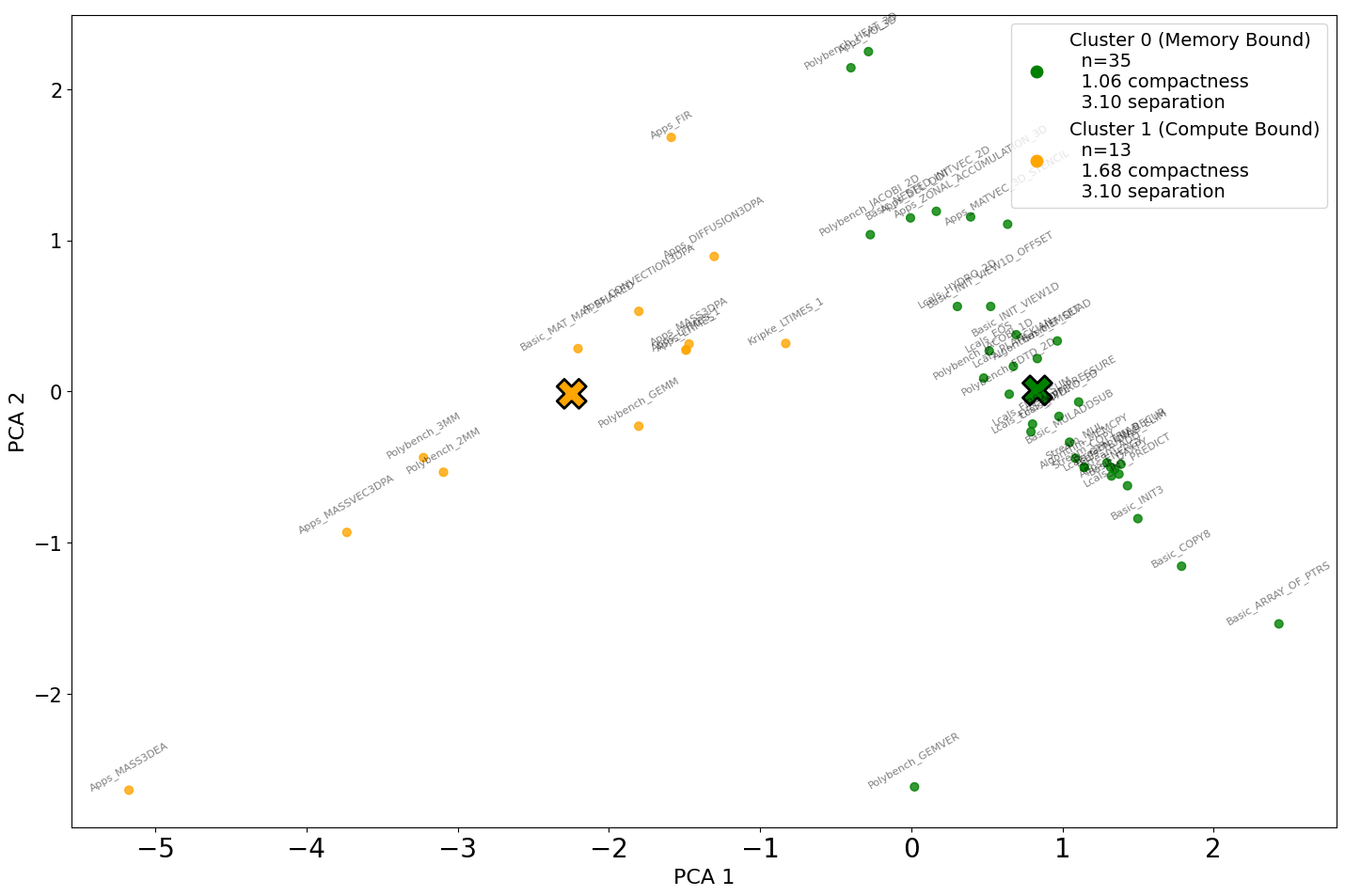}
        \caption{2D PCA visualization of $n$-dimensional \kmeans \ clustering}
        \label{fig:ncu_kmeans_scatter}
    \end{subfigure}
    \begin{subfigure}{\columnwidth}
        \centering
        \includegraphics[width=\columnwidth]{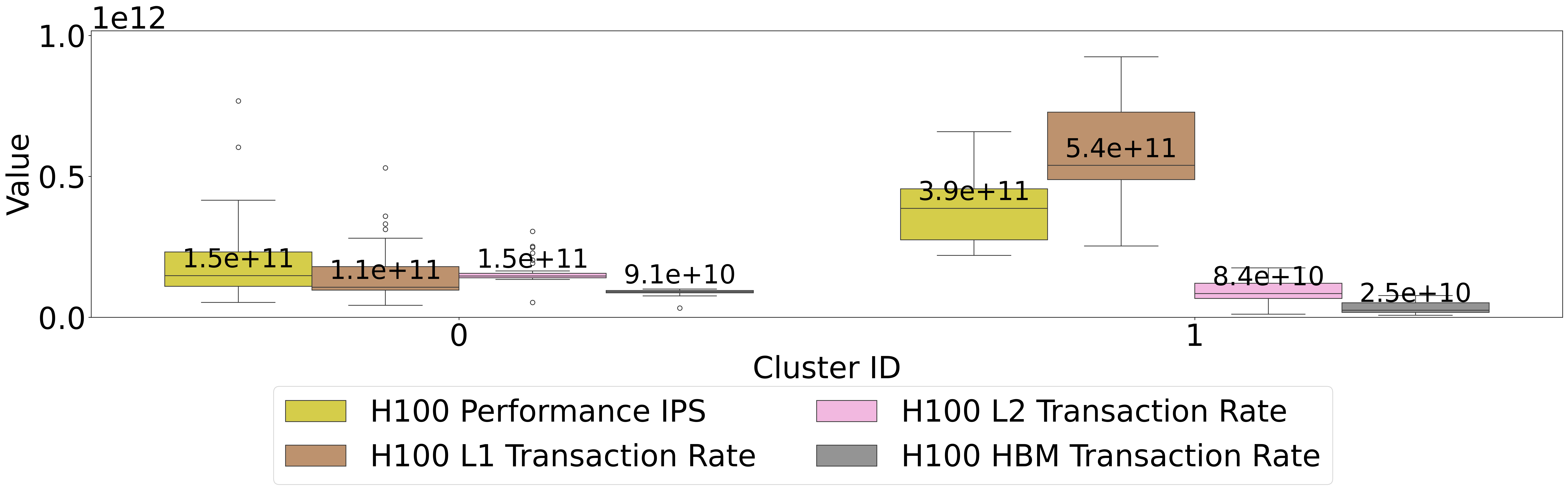}
        \caption{Distribution of metric values for each cluster}
        \label{fig:ncu_kmeans_barplot}
    \end{subfigure}
    \caption{K-means clustering using \ncu \ metrics}
    \label{fig:ncu_kmeans}
\end{figure}
\begin{figure}[htb]
    \centering
    \begin{subfigure}{\columnwidth}
        \centering
        \includegraphics[width=.925\columnwidth]{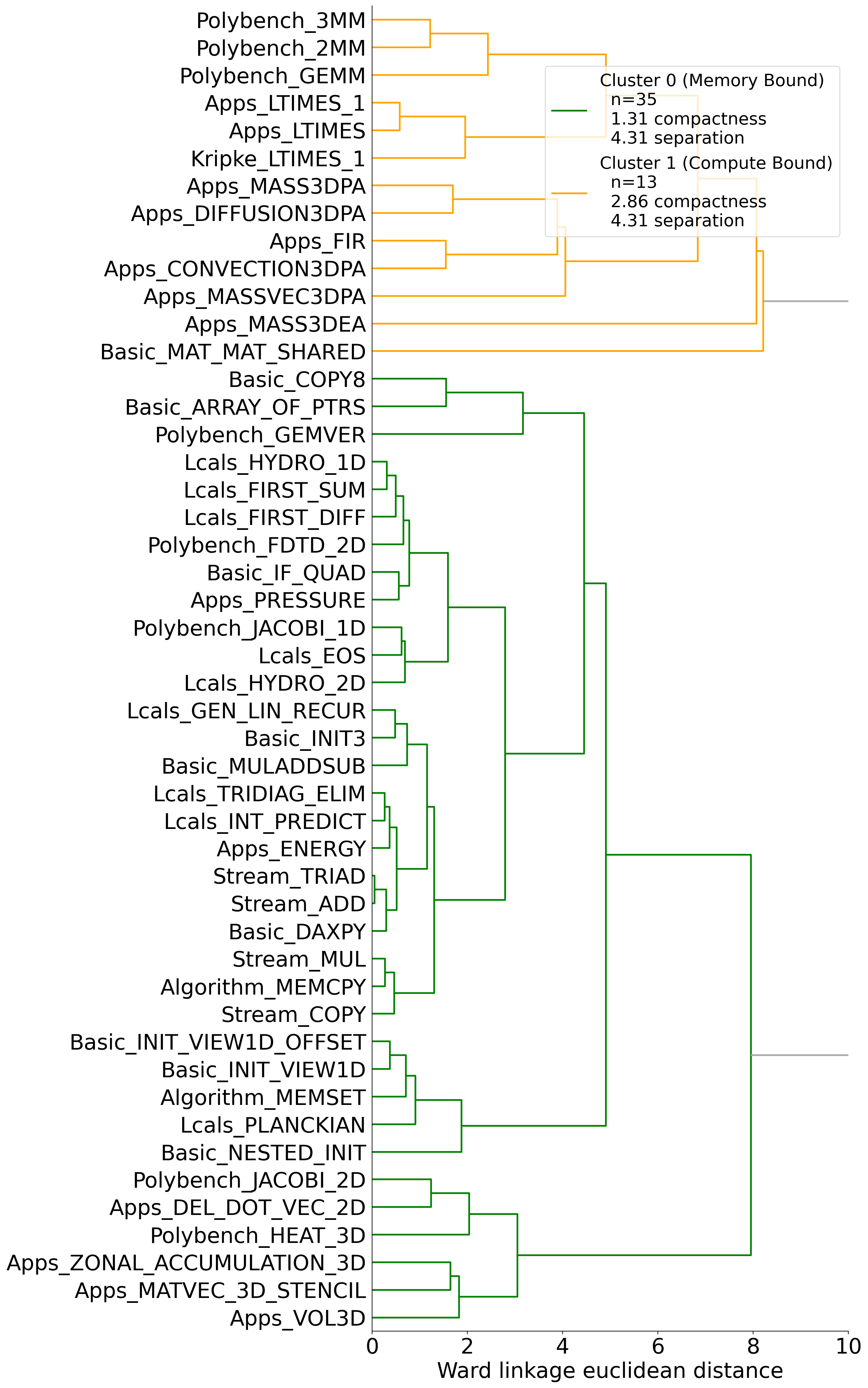}
        \caption{Dendrogram visualization of agglomerative clustering}
        \label{fig:ncu_topdown_agg_dendrogram}
    \end{subfigure}
    \begin{subfigure}{\columnwidth}
        \centering
        \includegraphics[width=\columnwidth]{figures/images/boxplot_clusters_topdown-nsight_SPR67108864NCU1073741824_ward_euclidean.png}
        \caption{Distribution of metric values for each cluster}
        \label{fig:ncu_topdown_agg_barplot}
    \end{subfigure}
    \caption{Agglomerative clustering using \topdown \ and \ncu \ metrics}
    \label{fig:ncu_topdown_agg}
\end{figure}
\begin{figure}[htb]
    \centering
    \begin{subfigure}{\columnwidth}
        \centering
        \includegraphics[width=\columnwidth]{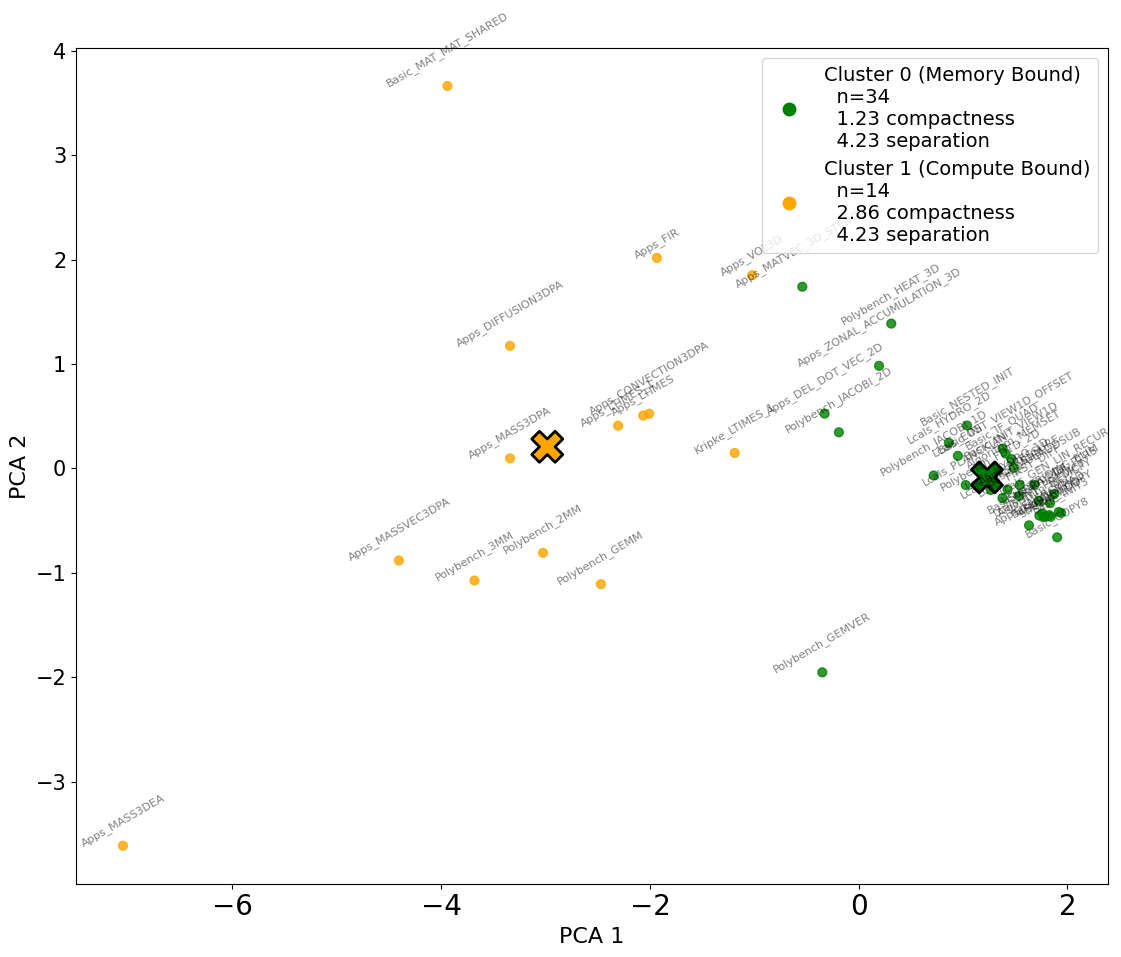}
        \caption{2D PCA visualization of $n$-dimensional \kmeans \ clustering}
        \label{fig:ncu_topdown_kmeans_scatter}
    \end{subfigure}
    \begin{subfigure}{\columnwidth}
        \centering
        \includegraphics[width=\columnwidth]{figures/images/boxplot_clusters_topdown-nsight_SPR67108864NCU1073741824_kmeans.png}
        \caption{Distribution of metric values for each cluster}
        \label{fig:ncu_topdown_kmeans_barplot}
    \end{subfigure}
    \caption{K-means clustering using \topdown \ and \ncu \ metrics}
    \label{fig:ncu_topdown_kmeans}
\end{figure}

\section{Performance Similarity of Kernels in the \rajaperf}
\label{sec:agglomerative_kmeans}

We evaluate performance similarity
using both \agg \ clustering and \kmeans \ clustering, and three combinations of data to understand per-platform and across-platform performance similarity. The three combinations of data are as follows: 1) only CPU hardware metrics, 2) only GPU hardware metrics, and 3) combined CPU and GPU  
hardware metrics.




\subsection{Performance Similarity on CPU}
\label{sec:cpu_agreement}

We first evaluate clustering on the CPU performance dataset.
Cluster 0 in Figure~\ref{fig:topdown_agg_dendrogram} and Figure~\ref{fig:topdown_kmeans_scatter} represent kernels that are memory bound, with a median runtime of 93.5\% (Figure~\ref{fig:topdown_agg_barplot}) and 93.2\% (Figure~\ref{fig:topdown_kmeans_barplot}) in the memory bound category.
Cluster 1 represents kernels which are compute bound, with a median core bound of 13.8\%, fetch bandwidth of 2.4\% and fetch latency of 0.6\% for \agg \ clustering and 13.9\%, 2.5\%, and 0.7\% for \kmeans \ clustering.

\subsubsection{Partitioning}
\label{sec:cpu_partitioning}

K-means classifies Apps\_Zonal\_Accumulation\_3D, Basic\_Array\_of\_Ptrs, and Polybench\_Jacobi\_2D as memory bound kernels, whereas agglomerative clustering classifies them as compute bound kernels.
Figure~\ref{fig:allkernels-raw-topdown} shows that these kernels cannot be placed into either category based on the 
\topdown\ metrics alone, as each of these kernels have values that match the characteristics from both clusters. 
This indicates that more performance information is required to correctly partition of kernels on the CPU.

\subsubsection{Compactness} 
\label{sec:cpu_compactness}
In Figures~\ref{fig:topdown_agg} and~\ref{fig:topdown_kmeans}, the memory bound cluster not only contains the most kernels, but it is also more compact than the compute bound cluster.
We observe from Table~\ref{table:compact_separate} that the memory bound cluster is up to 5.59$\times$ more compact than the compute bound cluster, indicating that the memory bound kernels are more similar than the compute bound kernels on the CPU.

\subsubsection{Separation}
\label{sec:cpu_separation}
We observe a higher separation for the \kmeans \ clusters, but tighter compactness for the \agg \ clusters.
K-means clustering classifies more kernels as memory bound than  \agg \ clustering. 
As a result, the memory bound cluster shows about $5\%$ higher separation than \agg \ clustering, while being less compact. 



\subsection{Performance Similarity on GPU}
\label{sec:gpu_agreement}

Because kernel performance characteristics may differ on the GPU, we  perform clustering on the GPU performance dataset by itself.
Cluster 0 in Figures~\ref{fig:ncu_agg_dendrogram} and~\ref{fig:ncu_kmeans_scatter} represents kernels that are memory bound, with a median HBM transaction rate of 9.1e10 transactions/sec and median L2 transaction rate of 1.5e11 transactions/sec (Figures~\ref{fig:ncu_agg_barplot} and~\ref{fig:ncu_kmeans_barplot}).
Cluster 1 represents kernels with high IPS and L1 transaction rate, with a median IPS of 3.9e11 and median L1 transaction rate of 5.4e11 transactions/sec.

\subsubsection{Partitioning}
\label{sec:gpu_partitioning}

The clustering methods produce identical clusters, \ie in Figure~\ref{fig:ncu_agg_dendrogram} we see the exact same clustering illustrated in Figure~\ref{fig:ncu_kmeans_scatter}. Therefore, the distribution of values in Figure~\ref{fig:ncu_agg_barplot} is also the same as Figure~\ref{fig:ncu_kmeans_barplot}.

\subsubsection{Compactness}
\label{sec:gpu_compactness}

From Table~\ref{table:compact_separate}, we denote that the memory bound cluster is $1.58\times$ more compact than the compute bound cluster, indicating that the similarity of the kernels in the memory bound cluster is higher.

\subsubsection{Separation}
\label{sec:gpu_separation}
Because the clusters are identical, separation between the clustering algorithms is also the same. However, we can compare across the datasets, and from Table~\ref{table:compact_separate} we observe that separation of the clusters is higher for the \ncu \ dataset than the \topdown \ dataset.

\begin{table*}[htb]
\centering
\small
\setlength{\tabcolsep}{3pt}
\caption{\label{table:compact_separate}For each cluster shown in the form (<cluster0>, <cluster1>), we show the number of kernels in each cluster as the result of the partitioning, the average per cluster compactness (lower is better), and the average cluster separation (higher is better). The ratio columns are derived by taking cluster 1 divided by cluster 0, and the relative difference columns is the division of the higher ratio divided by the smaller ratio. The datasets are standardized as described in Section~\ref{sec:hardware_metrics}.}
\vspace{-0.25cm}
\begin{tabular}{l | cc | ccccc | ccc}
\toprule
& \multicolumn{2}{c|}{Partitioning}
& \multicolumn{5}{c|}{Compactness (lower is better)}
& \multicolumn{3}{c}{Separation (higher is better)} \\
& Agglomerative & K-means
& Agglomerative & Ratio\ \ \  & K-means & Ratio & Relative 
& Agglomerative & K-means & Relative\ \ \ \\
\midrule
Top-down      & (28, 22) & (31, 19) & (0.34, 1.90) & 5.59$\times$ & (0.45, 1.94)  & 4.31$\times$ & 1.30$\times$ & 2.68 & 2.82 & 1.05$\times$ \\
\ncu           & (35, 13) & (35, 13) & (1.06, 1.68) & 1.58$\times$ & (1.06, 1.68)  & 1.58$\times$ & 1.00$\times$ & 3.10 & 3.10 & 1.00$\times$ \\
Top-down and \ncu  & (35, 13) & (34, 14) & (1.31, 2.86) &2.18$\times$ & (1.23, 2.86)  & 2.33$\times$ & 1.07$\times$ & 4.31 & 4.23 & 1.02$\times$ \\
\bottomrule
\end{tabular}
\vspace{-0.25cm}
\end{table*}

\subsection{Multi-Platform Performance Similarity}
\label{sec:cpu_gpu_agreement}

With consistent memory bound and compute bound patterns across the CPU and GPU metrics, we are able to combine the metrics into a single dataset and visualize similar performance characteristics.
In both Figure~\ref{fig:ncu_topdown_agg_dendrogram} and Figure~\ref{fig:ncu_topdown_kmeans_scatter}, cluster 0 represents kernels with a median memory bound of approximately 92\% as shown in Figures~\ref{fig:ncu_topdown_agg_barplot} and~\ref{fig:ncu_topdown_kmeans_barplot}, and a median HBM transaction rate of 9.1e10 transactions/sec. 
Cluster 1 represents the compute bound kernels with a median core bound of approximately 15\%, and median IPS rate of approximately 4e11.
As we expect, most of the memory bound and compute bound kernels cluster together, regardless of architecture. 
For example, in Figure~\ref{fig:ncu_topdown_agg_barplot}, memory bound kernels are 2.2$\times$ more memory bound on CPU and have a 3.6$\times$ higher HBM transaction rate on GPU.
The compute bound kernels have an average of 3.1$\times$ higher core bound on CPU, and 2.6$\times$ higher IPS on GPU.
This pattern indicates that for the \rajaperf, CPU/GPU memory and compute characteristics are heavily correlated.

\subsubsection{Partitioning}
\label{sec:cpu_gpu_partitioning}
\begin{figure*}[h!]
     \centering
     \begin{subfigure}{0.49\textwidth}
         \centering
         \includegraphics[width=0.95\textwidth]{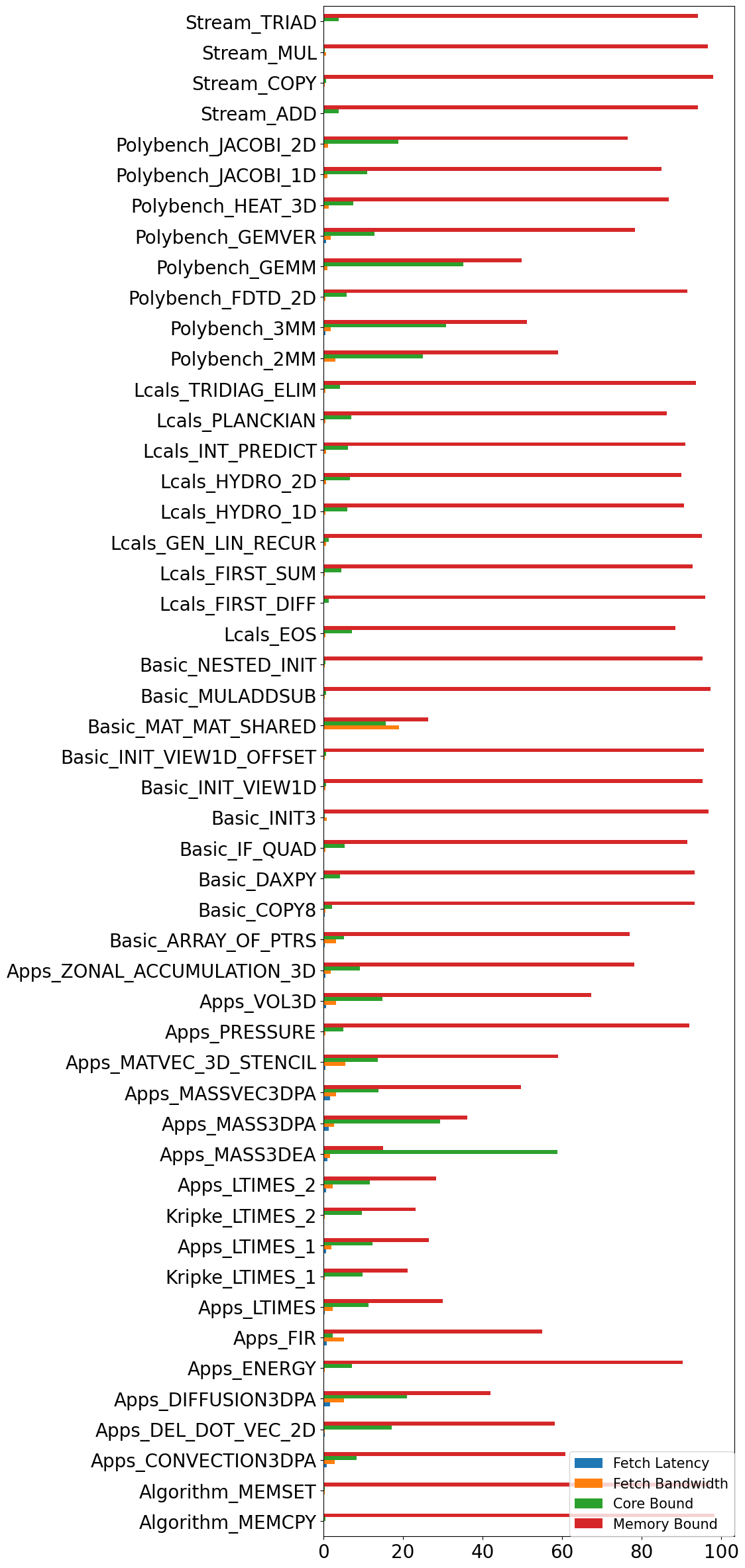}
                 \vspace{-0.2cm}
         \caption{\label{fig:allkernels-raw-topdown}Reference \topdown\ values for all kernels}
    \end{subfigure}
    \hfill
    \begin{subfigure}{0.49\textwidth}
        \centering
        \includegraphics[width=0.96\textwidth]{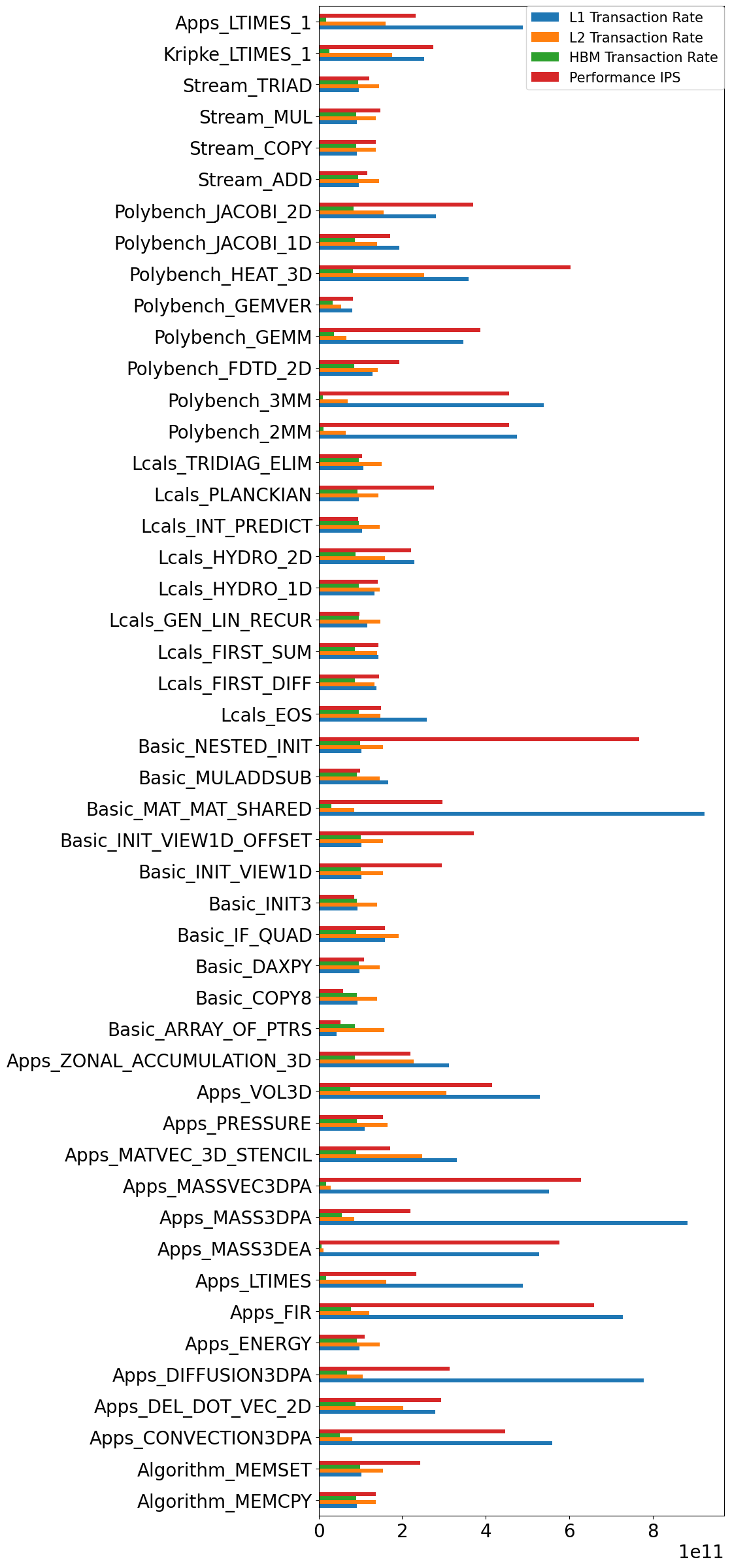}
        \vspace{-0.5cm}
        \caption{\label{fig:allkernels-raw-nsight}Reference \ncu\ values for all kernels}
    \end{subfigure}
    \caption{Reference CPU and GPU metric values for all kernels}
\end{figure*}

There is only one different kernel between the clustering methods, Apps\_VOL3D. 
Apps\_VOL3D is memory bound according to \agg \ clustering, whereas for \kmeans \ it is compute bound. 
This kernel has mostly compute bound characteristics on the CPU and both memory bound and compute bound characteristics on the GPU as shown in Figure~\ref{fig:allkernels-raw-topdown} and Figure~\ref{fig:allkernels-raw-nsight}. 
We expect cases like Apps\_VOL3D with significantly different performance characteristics on CPU and GPU to be in the middle of the two centroids, therefore difficult to cluster with only two centroids.
In both \agg \ and \kmeans \ clustering, the partitioning of the combined dataset is most similar to the \ncu \ dataset in Section~\ref{sec:gpu_agreement}.

\subsubsection{Compactness}
\label{sec:cpu_gpu_compactness}
Referring to the Compactness columns in Table~\ref{table:compact_separate}, the memory bound cluster is as much as 2.33$\times$ more compact and contains more kernels than the compute bound cluster.
We also visualize the compactness in the boxplots (\eg Figure~\ref{fig:ncu_topdown_agg_barplot}), where across all three datasets the inter-quartile range for the memory bound cluster is always significantly smaller.
Since the memory bound cluster is more compact in all three datasets, we can conclude that the kernels in the memory bound cluster perform more similarly to each other than the kernels in the compute bound cluster.

\subsubsection{Separation}
\label{sec:cpu_gpu_separation}
Looking at the Separation columns in Table~\ref{table:compact_separate}, we find the \ncu \ dataset has higher separation than the \topdown\ dataset. 
However, the separation is the highest when combining \topdown \ and \ncu \ (4.31), at the tradeoff of the memory bound cluster being least compact out of the three datasets (see 1.31 in the Compactness columns).
Thus, the performance characteristics between the clusters are most different for the combined dataset, but the clusters become less similar because there is additional information for each kernel.


\begin{figure*}[htb]
    \centering
    \begin{subfigure}[t]{\columnwidth}
        \centering
        \includegraphics[width=0.92\columnwidth]{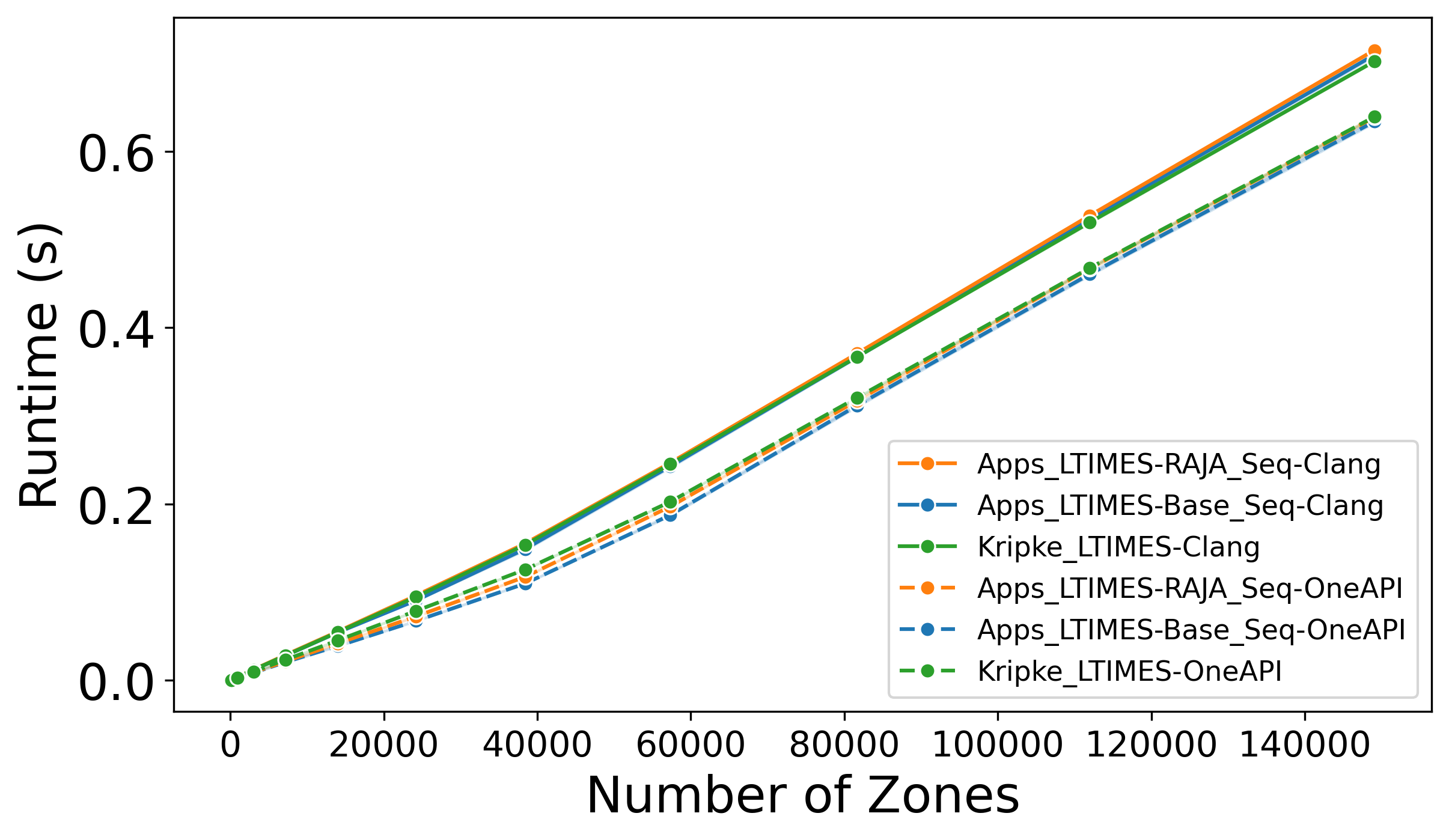}
        \caption{\label{fig:runtime-ltimes-cpu}Intel Sapphire Rapids CPU. RAJA\_Seq and Base\_Seq (C++) implementations from \rajaperfabbrev. Clang, Intel OneAPI compilers.}
    \end{subfigure}
    \hfill
    \begin{subfigure}[t]{\columnwidth}
        \centering
        \includegraphics[width=0.92\columnwidth]{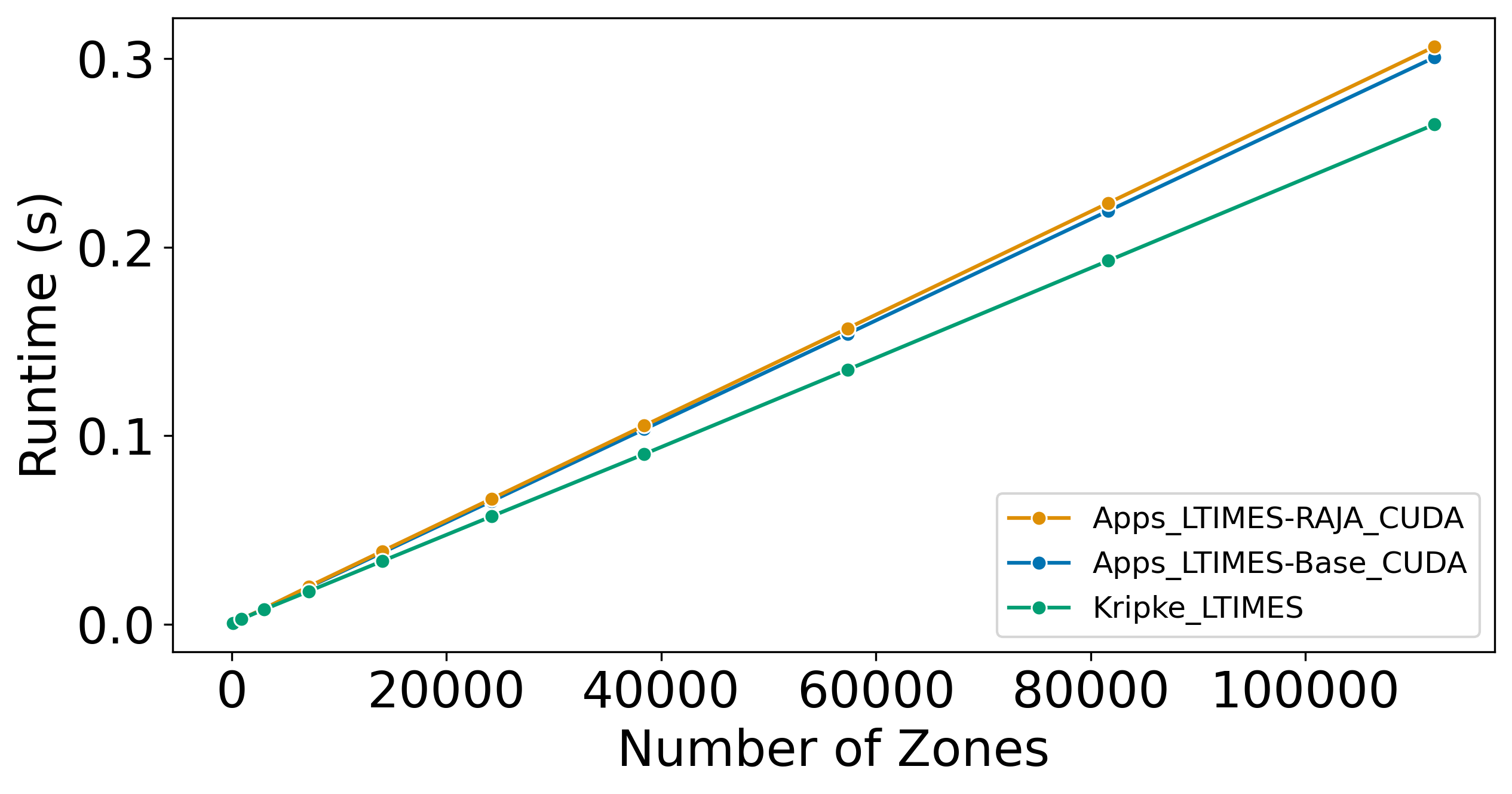}
        \caption{\label{fig:runtime-ltimes-gpu}NVIDIA H100 GPU.  From \rajaperfabbrev, showing RAJA\_CUDA and Base\_CUDA implementations.
        }
    \end{subfigure}
    \caption{\label{fig:ltimes}LTIMES kernel runtime: Kripke implementation vs. \rajaperf \ implementations.}
\end{figure*}
\begin{table*}[]
\centering
\small
\setlength{\tabcolsep}{3pt}
\caption{\label{table:rltimes}Average Euclidean distance from the Apps\_LTIMES kernel in the \rajaperf \ to other kernels. The phrase "all sizes" denotes where the average distance is shown across multiple samples.}
\vspace{-0.25cm}
\begin{tabular}{l c c c c c l}
\toprule
& Apps\_LTIMES & Kripke\_LTIMES & All LTIMES & Other closest & Relative other closest & Other closest \\
& all sizes & all sizes & all sizes & kernel (single) &  vs any LTIMES & kernel name \\
\midrule
\topdown      & 0.58 & 1.24 & 0.91 & 1.43 & 1.57$\times$ & Apps\_DEL\_DOT\_VEC\_2D\\
\ncu           & 0.004 & 1.02 & 0.51 & 1.67 & 3.27$\times$ & Basic\_MAT\_MAT\_SHARED \\
\topdown\ \& \ncu  & 0.58 & 1.57 & 1.08 & 2.76 & 2.56$\times$ & Polybench\_2MM \\
\bottomrule
\vspace{-0.25cm}
\end{tabular}
\end{table*}




\section{Similarity of Multiple Implementations of the LTIMES kernel}
\label{sec:kripke_vs_rajaperf}


To verify correctness of partitioning, we introduce labeled examples. 
We demonstrate the LTIMES kernel as one labeled example, as LTIMES is 
implemented in both the Kripke proxy application and in \rajaperfabbrev. 
We begin by manually verifying the source code implementations of the LTIMES kernel in both benchmarks to be equivalent.
Figure~\ref{fig:ltimes} shows the execution time of the LTIMES kernel implemented in both Kripke and \rajaperfabbrev \ for
a CPU system in Figure~\ref{fig:runtime-ltimes-cpu} and for 
a GPU system in Figure~\ref{fig:runtime-ltimes-gpu},
varying the kernel size parameter (number of zones).
For LTIMES in \rajaperfabbrev, we show a RAJA implementation and Base programming model implementation (C++ on CPU, CUDA on GPU).
On the CPU, we also compare binaries generated by two compilers, 
Clang 19.1.3 and Intel OneAPI 2025.2.0. 
While OneAPI generates a more performant binary than Clang, 
the runtime difference between LTIMES in Kripke and LTIMES in \rajaperfabbrev \ is 
within 1\% for both compilers, for both RAJA and Base implementations. 
On the GPU, Kriple LTIMES is consistently 15\% faster than
the LTIMES kernel in \rajaperfabbrev. 


We further evaluate the similarity between the \altimes \ and \kltimes \ kernels using the hardware metrics (Section~\ref{sec:hardware_metrics}).
Table~\ref{table:rltimes} details which three kernels are the most similar to \altimes\ on each dataset, with a lower distance indicating higher similarity. The possible options are: (1) \altimes\ ran at a different problem size (\altimes\_X), (2) \kltimes, or (3) any other kernel in the \rajaperf.
First, we notice the distance for \altimes\ on the \ncu\ dataset is significantly smaller than the \topdown\ distance. 
For the \ncu \ dataset, the chosen problem size for \altimes\_1 is only $1.012\times$ larger than Apps\_LTIMES, thus the kernels are very similar (0.004).
On the \topdown \ set, the problem size is $5.323\times$ larger for \altimes\_2, therefore the similarity is significantly lower (0.58). 
However, despite the problem size differences we see these executions of \altimes\ are the most similar, and therefore we observe that source code has the largest impact on kernel similarity.

\kltimes \ is always the most similar non-\altimes \ kernel on all of the datasets. 
Based on our similarity distance metric, the average similarity of the LTIMES kernels is from 1.57-3.27$\times$ greater than any other kernel.
Furthermore, we observe that the similarity of the closest non-LTIMES kernel increases significantly (2.76) when combining the CPU and GPU datasets.
The geometric mean of the relative similarity values on each dataset (1.57$\times$ and 3.27$\times$) is 2.27$\times$, so the 2.56$\times$ relative similarity on the combined multi-platform dataset indicates that combining the datasets improves the similarity measure.
\kltimes \ being significantly more similar to \altimes \ than the next closest kernel 
in all three datasets indicates that we are able to distinguish the LTIMES kernel based on performance data.

\begin{figure*}[htb]
\centering
\setlength{\tabcolsep}{2pt}

\begin{tabular}{c c c c c c c}
    \raisebox{8.9cm}[0pt][0pt]{%
      \rotatebox{90}{%
        \parbox{2.5cm}{\centering
          Average silhouette width (higher is better)
        }%
      }%
    } &
    \begin{subfigure}{0.14\textwidth}
        \includegraphics[width=\linewidth]{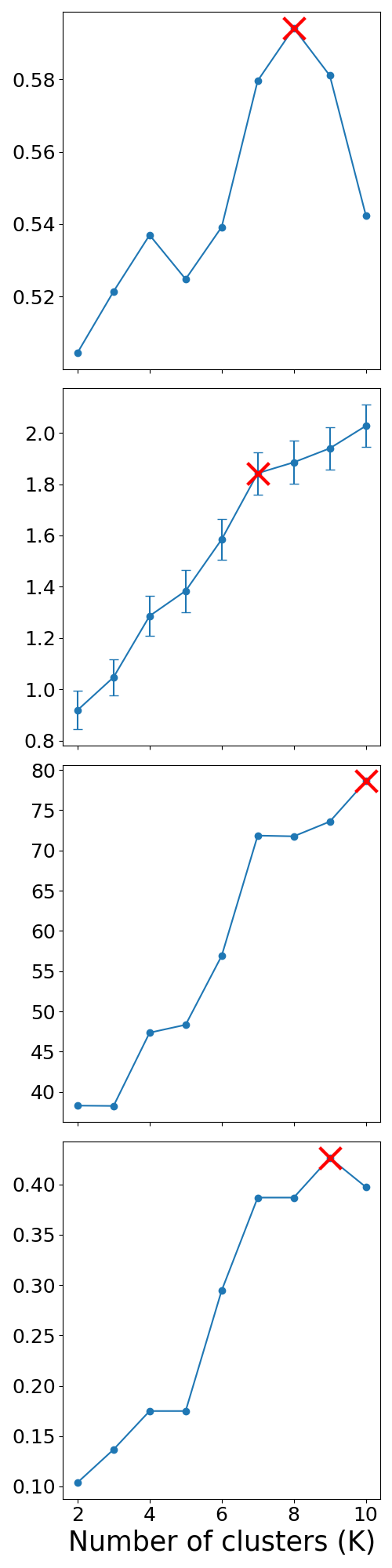}
        \caption{Agglom. clustering, Top-down metrics}\label{fig:selection-agg-topdown}
    \end{subfigure} &
    \begin{subfigure}{0.14\textwidth}
        \includegraphics[width=\linewidth]{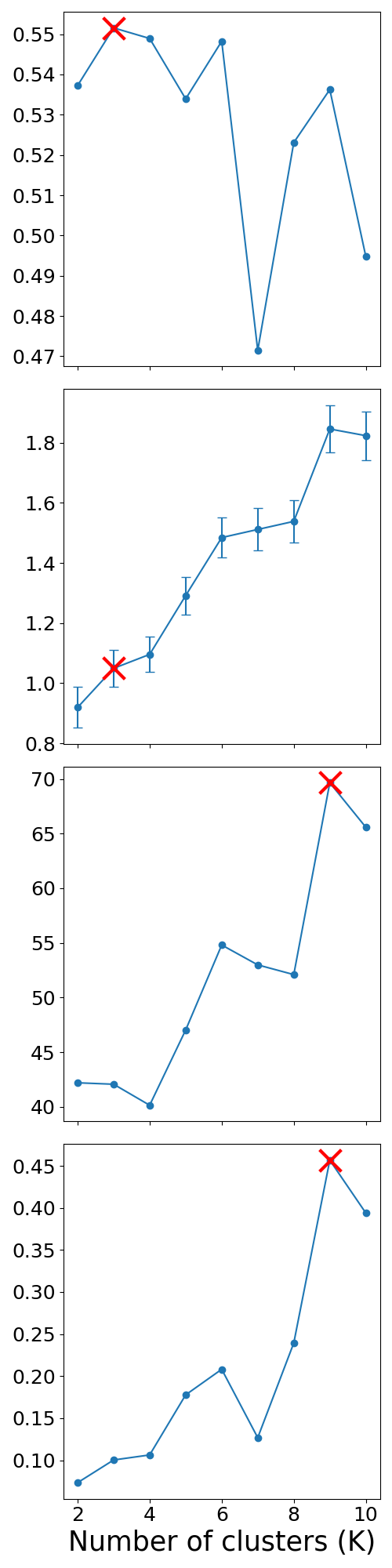}
        \caption{K-means clustering, Top-down metrics}\label{fig:selection-kmeans-topdown}
    \end{subfigure} &
    \begin{subfigure}{0.14\textwidth}
        \includegraphics[width=\linewidth]{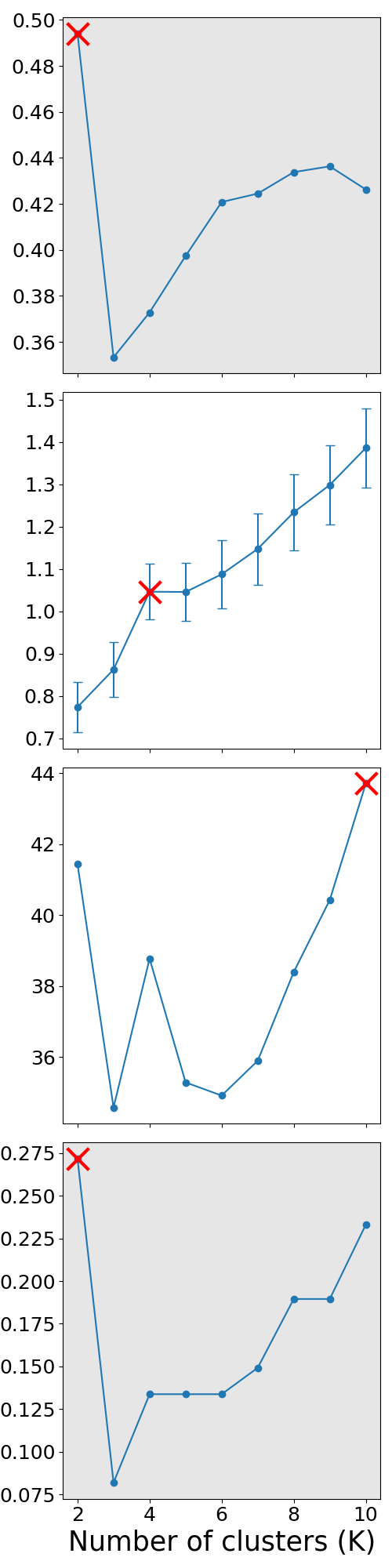}
        \caption{Agglom. clustering, NCU metrics\\}\label{fig:selection-agg-ncu}
    \end{subfigure} &
    \begin{subfigure}{0.14\textwidth}
        \includegraphics[width=\linewidth]{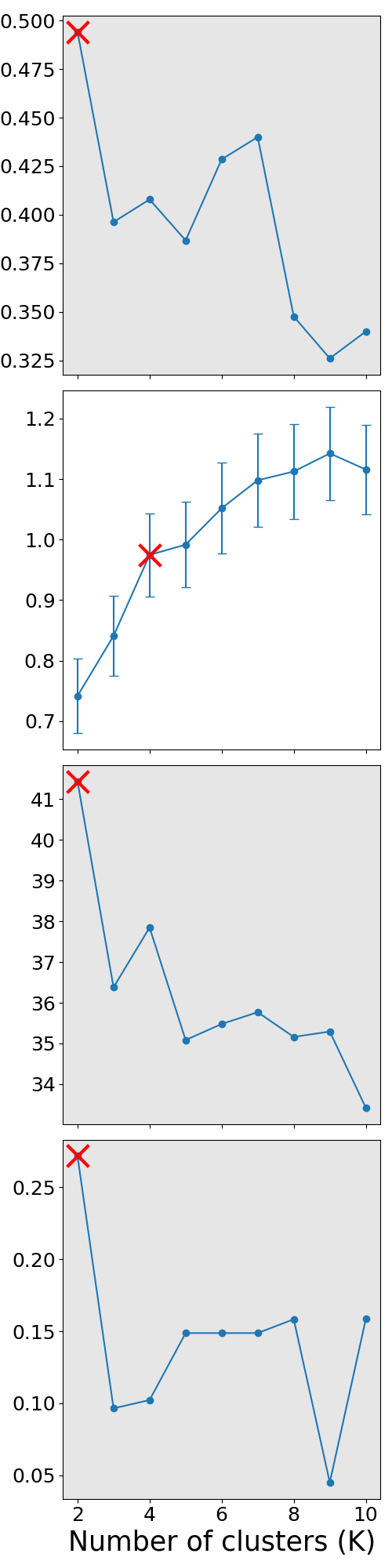}
        \caption{K-means clustering, NCU metrics\\}\label{fig:selection-kmeans-ncu}
    \end{subfigure} &
    \begin{subfigure}{0.14\textwidth}
        \includegraphics[width=\linewidth]{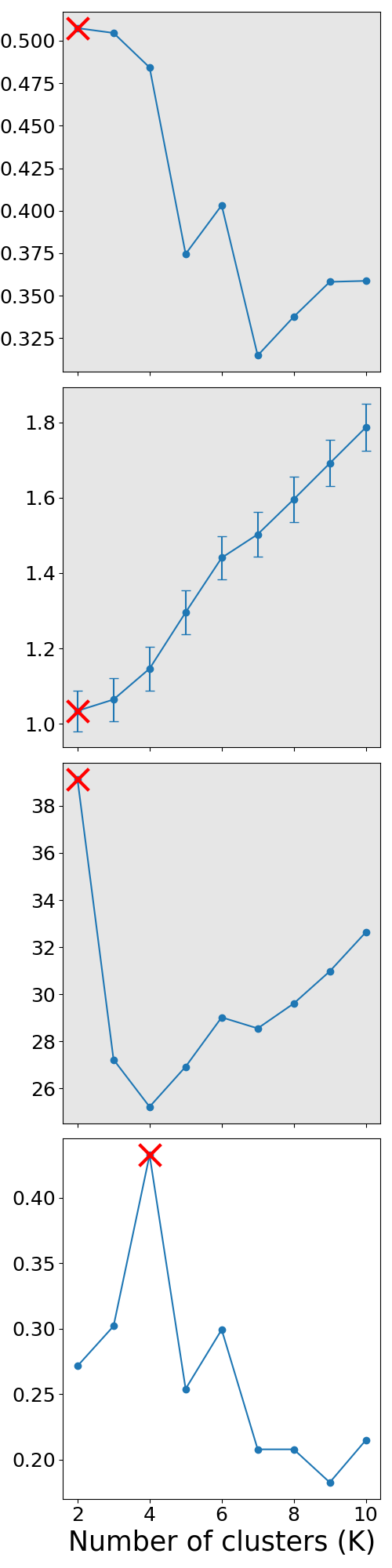}
        \caption{Agglom. clustering, Top-down + NCU}
    \end{subfigure} &
    \begin{subfigure}{0.14\textwidth}
        \includegraphics[width=\linewidth]{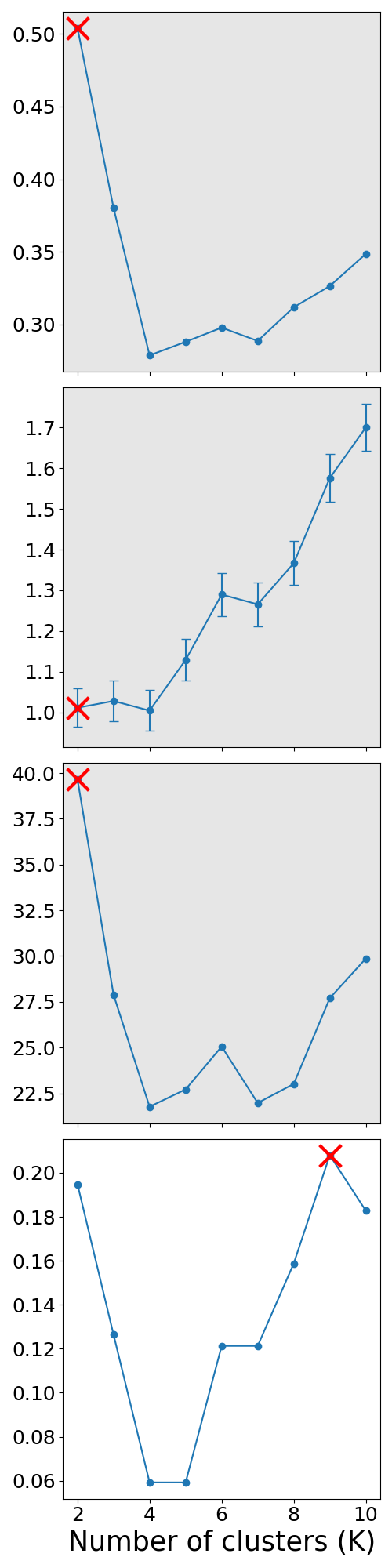}
        \caption{K-means clustering, Top-down + NCU}
    \end{subfigure} \\

        \raisebox{6.9cm}[0pt][0pt]{%
      \rotatebox{90}{%
        \parbox{2.5cm}{\centering
          Gap Statistic (Tibshirani rule)
        }%
      }%
    } &
    & & & & & \\

        \raisebox{4.7cm}[0pt][0pt]{%
      \rotatebox{90}{%
        \parbox{2.5cm}{\centering
          Variance Ratio Criterion (higher is better)
        }%
      }%
    } &
    & & & & & \\

        \raisebox{2.25cm}[0pt][0pt]{%
      \rotatebox{90}{%
        \parbox{2.5cm}{\centering
          Dunn Index (higher is better)
        }%
      }%
    } &
    & & & & & \\
\end{tabular}
\vspace{-1.5cm}
\caption{
Comparison of \agg \ and \kmeans \ clustering using different metrics and selection criteria.
    Rows are selection criteria (from top to bottom): 1) average silhouette width, 2) gap statistic with Tibshirani stopping rule, 3) Calinski-Harabasz variance-ratio criterion, and 4) Dunn index. The red "X" indicates which $k$, that is the number of clusters, the selection method determined is optimal. Grayscale plots designate cases where $k=2$.
}

\label{fig:selection_matrix}
\end{figure*}

\section{Selecting the Number of Clusters}
\label{sec:selection}

Selecting the optimal number of clusters is a long standing problem~\cite{10.1145/3606274.3606278},
as both higher separation and lower compactness are desired, however they are often not jointly optimizable.
We evaluated six different cluster selection methods surveyed in the literature~\cite{10.1145/3606274.3606278}.
These are: 1) average silhouette width, 2) gap statistic with Tibshirani stopping rule, 
3) Calinski-Harabasz variance-ratio criterion, and 4) Dunn Index.
Additionally, we evaluated the Davies-Bouldin Index~\cite{davies}, and Bayesian Information Criterion~\cite{Pelleg2000XmeansEK}, 
but omit them based on experimental results, as neither method selected two clusters for any dataset.

Two of these methods (Silhouette and Dunn) are distance-based methods, meaning the quality of the 
clustering is determined by an equation involving Euclidean distance.
The silhouette width measure~\cite{ROUSSEEUW198753} compares the average distance of a point against each cluster center, with a higher silhouette width indicating a point is much closer to its own cluster than other clusters.
The Dunn Index~\cite{Dunn01011973} compares cluster diameter against cluster separation as a ratio of the smallest cluster separation to the largest cluster diameter, with a higher score indicating maximal smallest-cluster separation and minimal largest-cluster diameter.

Additionally, we use the simulation-based gap statistic with Tibshirani stopping rule~\cite{https://doi.org/10.1111/1467-9868.00293}, an algorithm clustering uniformly random datasets and compares the sum of squared errors from clustering via the random dataset against the target dataset. We apply this method for both \kmeans \ (via \kmeans \ simulated clusters) and \agg \ clustering (via \agg \ simulated clusters). The stopping criteria is dependent on the gap value of $k$ being less than the gap value of $k-1$ minus the standard deviation of the $k-1$ estimates.
Finally, the variance ratio criterion~\cite{Calinski01011974} is computed from the ratio of the between-group sum of squares (BGSS) over the within-group sum of squares (WGSS). Maximizing BGSS increases separation and minimzing WGSS increases cluster compactness, so a higher VRC score is better.

In Figure~\ref{fig:selection_matrix}, we illustrate the result of applying four cluster selection methods on different combinations of the three datasets (\topdown, \ncu, and \topdown \ + \ncu) and two clustering methods (\agg \ and \kmeans). 
We determine optimal $k$ value is $k=2$ from several factors.
One reason we choose $k=2$ not only because it is by consensus the most common value of $k$ in Figure~\ref{fig:selection_matrix}, shown as grayscale plots. 
Furthermore, we find $k=2$ results in the most interpretable clusters, and the property of a consistent $k$ value makes the results comparable across datasets and clustering methods.
From the dataset of \topdown \ and \ncu \ metrics, we find there are two performance characteristics: (1) memory bound kernels and (2) compute bound kernels (see Section~\ref{sec:agglomerative_kmeans}).
Figure~\ref{fig:selection_matrix} indicates we would need an increased set of metrics to define additional clusters with interpretable performance characteristics.

From Figure~\ref{fig:selection_matrix}, we also observe how separability impacts cluster selection.
In Table~\ref{table:compact_separate}, we note that the \topdown+\ncu \ dataset has the highest separability, followed by the \ncu \ dataset, and then the \topdown \ dataset.
Interestingly when combining the \topdown \ and \ncu \ data into one dataset, the highest amount of selection methods (six out of eight) choose $k=2$ as optimal $k$ value.
The \ncu \ dataset is also highly separable at two clusters, as in five out of eight cases the best choice is $k=2$.
Finally, for the \topdown \ dataset, no selection methods choose $k=2$ for the \topdown \ dataset, instead favoring higher $k$ values.
\section{Related Work}
\label{sec:related}

In previous work, we have taken an in depth look into the \rajaperf ~\cite{llnlRAJAPerf2024} and analyzed the performance of \rajaperfabbrev \ kernels at a high-level to identify key bottlenecks using hierarchical clustering on the set of \topdown \ metrics.
We also analyzed \rajaperfabbrev \ across seven different hardware architectures, 
including evaluation of the \ncu \ metrics~\cite{10.1145/3754598.3754668}.
Current work addresses selection of the \topdown \ metrics which are indicators of performance, 
excludes irrelevant or misleading metrics, and  evaluates clustering methods thoroughly (\ie via selection methods). 
Additionally, we include \ncu \ metrics in our analysis, which significantly improves 
assignment of the memory bound and compute bound clusters.


Evaluating code similarity is an active research area.
The ECP Proxy App Project~\cite{richards:ecp2021} studied the ability to assess performance similarity across 
eight proxy/parent combinations on Intel Skylake and IBM Power9 CPU architectures, and two proxy/parent combinations on NVIDIA V100.
Leveraging LDMS~\cite{LDMS} to run experiments, approximately 700 hardware counters were collected using PAPI~\cite{papi} on the CPUs, and approximately 100 hardware counters on the GPU.
An explicit effort was made to avoid hand-picking metrics to minimize bias.
The study leveraged the cosine similarity metric, as they found the distance-based similarity metrics 
had low interpretability on their dataset, and the angle-based cosine similarity resulted in more intuitive clusters.
The main finding between the two CPU architectures was that six out of eight proxy applications were most similar to their parent application, with two proxies miniQMC and XSBench not matching the performance of their parent.
On GPU, neither of the proxies (ExaMiniMD and sw4lite) clustered with their parent, and the authors concluded that further work was required to determine if the implementations are fundamentally different on the GPU.
Although our application pair (Kripke and RAJAPerf) is not proxy/parent, we are able to successfully demonstrate similarity between kernel implementations, and we provide a methodology that enables cross-platform similarity instead of being constrained to analysis on either the CPU or the GPU. 
Additionally, we found that the trend-focused cosine similarity metric, while useful for identifying patterns, 
would incorrectly classify certain derived metrics such as the Intel top-down percentage values. Therefore, we leverage Euclidean distance as a similarity metric instead.
Our study also targets a hand-picked set of metrics derived from multiple hardware counters, 
instead of using many raw hardware metrics, as noisy metrics can significantly affect the clustering results.

Attempts to characterize several benchmarks based on architecture-neutral spatial and temporal memory locality metrics
have not focused on clustering and similarity, but instead highlighted the differences in memory characterization 
between the different benchmarks~\cite{f0f3240e4a9510cc07515899834b8e7b122f1a38}.
Other work involves matching execution signatures
via a prediction framework that aims to
predict the performance of application phases by the performance of reference kernels~\cite{matching-signatures}. 
The framework was tested on three applications, with performance prediction error in the range of 0.4-18.7\%. 
The metrics used were raw counters and the goal focused on predicting kernel performance using a distance based 
similarity metric. Our work focuses on evaluating the similarity of computational kernels with known performance characteristics.



\section{Conclusions}
\label{sec:conclusions}

This work presents a practical approach to evaluate whether benchmarks are representative of simulation codes by comparing 
the performance of a computational kernel from two applications on two different platforms. 
By defining and validating performance similarity metrics, we demonstrate that similar kernels 
can be identified based on their performance characteristics. 
We use kernels from the Kripke proxy application and the \rajaperf \ as case studies for our proposed 
methodology, a more scalable framework for evaluating if benchmarks are representative of our 
HPC codes in the rapidly changing HPC hardware landscape.
In addition, we leverage a quantitative methodology for selecting the number of 
clusters alongside qualitative analysis.
Our focus on cluster interpretability enables the characterization 
of two categories of kernels with distinct performance characteristics.
Our proposed similarity metrics enable assessment of the similarity of computational 
kernels in our codes and the proxy applications we use to represent the codes.


\begin{acks}
This work was performed under the auspices of the U.S.~Department of Energy by Lawrence Livermore National Laboratory under Contract DE-AC52-07NA27344 and was supported by the LLNL-LDRD Program under Project No. 26-ERD-041 (LLNL-CONF-2018122). 
\end{acks}


%
%
%
%


\bibliographystyle{ACM-Reference-Format}

\bibliography{cite}

@inproceedings{10.1145/rajaperf-2026, author = {Abouelmagd, Amr and Boehme, David and Brink, Stephanie and Burmark, Jason and McKinsey, Michael and Skjellum, Anthony and Pearce, Olga}, title = {GPU Partitioning, Power, and Performance of the AMD MI300A}, year = {2026}, isbn = {9798400720673}, publisher = {Association for Computing Machinery}, address = {New York, NY, USA}, url = {https://doi.org/10.1145/3773656.3773680}, doi = {10.1145/3773656.3773680}, booktitle = {Proceedings of the Supercomputing Asia and International Conference on High Performance Computing in Asia Pacific Region}, pages = {215–227}, numpages = {13}, location = { }, series = {SCA/HPCAsia '26} }

@article{ROUSSEEUW198753,
title = {Silhouettes: A graphical aid to the interpretation and validation of cluster analysis},
journal = {Journal of Computational and Applied Mathematics},
volume = {20},
pages = {53-65},
year = {1987},
issn = {0377-0427},
doi = {https://doi.org/10.1016/0377-0427(87)90125-7},
url = {https://www.sciencedirect.com/science/article/pii/0377042787901257},
author = {Peter J. Rousseeuw}
}

@article{https://doi.org/10.1111/1467-9868.00293,
author = {Tibshirani, Robert and Walther, Guenther and Hastie, Trevor},
title = {Estimating the number of clusters in a data set via the gap statistic},
journal = {Journal of the Royal Statistical Society: Series B (Statistical Methodology)},
volume = {63},
number = {2},
pages = {411-423},
doi = {https://doi.org/10.1111/1467-9868.00293},
url = {https://rss.onlinelibrary.wiley.com/doi/abs/10.1111/1467-9868.00293},
eprint = {https://rss.onlinelibrary.wiley.com/doi/pdf/10.1111/1467-9868.00293},
year = {2001}
}

@article{Calinski01011974,
author = {T. Caliński and J Harabasz},
title = {A dendrite method for cluster analysis},
journal = {Communications in Statistics},
volume = {3},
number = {1},
pages = {1--27},
year = {1974},
publisher = {Taylor \& Francis},
doi = {10.1080/03610927408827101},
URL = { 
        https://doi.org/10.1080/03610927408827101
},
eprint = { 
        https://doi.org/10.1080/03610927408827101
}
}

@article{davies,
author = {Davies, David and Bouldin, Don},
year = {1979},
month = {05},
pages = {224 - 227},
title = {A Cluster Separation Measure},
volume = {PAMI-1},
journal = {Pattern Analysis and Machine Intelligence, IEEE Transactions on},
doi = {10.1109/TPAMI.1979.4766909}
}

@article{Dunn01011973,
author = {J. C. Dunn},
title = {A Fuzzy Relative of the ISODATA Process and Its Use in Detecting Compact Well-Separated Clusters},
journal = {Journal of Cybernetics},
volume = {3},
number = {3},
pages = {32--57},
year = {1973},
publisher = {Taylor \& Francis},
doi = {10.1080/01969727308546046},
URL = { 
        https://doi.org/10.1080/01969727308546046
},
eprint = { 
        https://doi.org/10.1080/01969727308546046
}
}

@inproceedings{Pelleg2000XmeansEK,
  title={X-means: Extending K-means with Efficient Estimation of the Number of Clusters},
  author={Dan Pelleg and Andrew W. Moore},
  booktitle={International Conference on Machine Learning},
  year={2000},
  url={https://api.semanticscholar.org/CorpusID:11243672}
}

@article{10.1145/3606274.3606278,
author = {Schubert, Erich},
title = {Stop using the elbow criterion for k-means and how to choose the number of clusters instead},
year = {2023},
issue_date = {June 2023},
publisher = {Association for Computing Machinery},
address = {New York, NY, USA},
volume = {25},
number = {1},
issn = {1931-0145},
url = {https://doi.org/10.1145/3606274.3606278},
doi = {10.1145/3606274.3606278},
journal = {SIGKDD Explor. Newsl.},
month = jul,
pages = {36–42},
numpages = {7}
}

@ARTICLE{kmeans,
author={Lloyd, S.},
journal={IEEE Transactions on Information Theory},
title={{Least squares quantization in PCM}},
year={1982},
volume={28},
number={2},
pages={129-137},
doi={10.1109/TIT.1982.1056489}
}

@inproceedings{10.1145/3754598.3754668,
author = {Yokelson, Dewi and Brink, Stephanie and Burmark, Jason and McKinsey, Michael and Bogale, Befikir and Lumsden, Ian and Taufer, Michela and Scogland, Tom and Pearce, Olga},
title = {Cross-Architecture Performance Analysis Using the RAJA Performance Suite},
year = {2025},
isbn = {9798400720741},
publisher = {Association for Computing Machinery},
address = {New York, NY, USA},
url = {https://doi.org/10.1145/3754598.3754668},
doi = {10.1145/3754598.3754668},
booktitle = {Proceedings of the 54th International Conference on Parallel Processing},
pages = {617–626},
numpages = {10},
location = {
},
series = {ICPP '25}
}

@inproceedings{llnlRAJAPerf2024,
author = {Pearce, Olga and Burmark, Jason and Hornung, Rich and Bogale, Befikir and Lumsden, Ian and McKinsey, Michael and Yokelson, Dewi and Boehme, David and Brink, Stephanie and Taufer, Michela and Scogland, Tom},
year = {2024},
month = {11},
pages = {1206-1218},
title = {RAJA Performance Suite: Performance Portability Analysis with Caliper and Thicket},
journal={2024 IEEE/ACM International Workshop on Performance, Portability and Productivity in HPC (P3HPC)},
doi = {10.1109/SCW63240.2024.00162}
}

@misc{benchpark,
  title = {{{Benchpark}}},
  author = {{LLNL}},
  year = {2023},
  copyright = {Apache 2.0 w/ LLVM Exception license},
  howpublished = {\url{https://github.com/LLNL/benchpark}}
}

@misc{kripke,
  title = {{{Kripke}}},
  author = {{LLNL}},
  year = {2015},
  month = jul,
  howpublished = {\url{http://github.com/LLNL/kripke}},
}

@techreport{kunen2015kripke,
  author       = {A. J. Kunen and T. S. Bailey and P. N. Brown},
  title        = {{KRIPKE-A Massively Parallel Transport Mini-App}},
  institution  = {Lawrence Livermore National Laboratory (LLNL)},
  address      = {Livermore, CA},
  year         = {2015},
  type         = {Tech. Rep.}
}

@inproceedings{hpctests2023-benchpark, 
   author = {Pearce, Olga and Scott, Alec and Becker, Gregory and Haque, Riyaz and Hanford, Nathan and Brink, Stephanie and Jacobsen, Doug and Poxon, Heidi and Domke, Jens and Gamblin, Todd}, 
   title = {{Towards Collaborative Continuous Benchmarking for {HPC}}}, 
   year = {2023}, 
   isbn = {9798400707858}, 
   publisher = {Association for Computing Machinery}, 
   address = {New York, NY, USA}, 
   url = {https://doi.org/10.1145/3624062.3624135}, 
   doi = {10.1145/3624062.3624135}, 
   booktitle = {Proceedings of the SC '23 Workshops of The International Conference on High Performance Computing, Network, Storage, and Analysis}, 
   pages = {627–635}, 
   numpages = {9}, 
   location = {Denver, CO, USA}, 
   series = {SC-W '23} 
}

@misc{ramble,
  title = {{{Ramble}}},
  author = {{Google}},
  year = {2023},
  copyright = {Apache-2.0 OR MIT},
  howpublished = {\url{https://github.com/GoogleCloudPlatform/ramble}}
}

@inproceedings{10.1007/978-3-031-97635-3_39,
author = {Bogale, Befikir and Lumsden, Ian and Sukkari, Dalal and Yokelson, Dewi and Brink, Stephanie and Pearce, Olga and Taufer, Michela},
title = {Surrogate Models for Analyzing Performance Behavior of HPC Applications Using the RAJA Performance Suite},
year = {2025},
isbn = {978-3-031-97634-6},
publisher = {Springer-Verlag},
address = {Berlin, Heidelberg},
url = {https://doi.org/10.1007/978-3-031-97635-3_39},
doi = {10.1007/978-3-031-97635-3_39},
booktitle = {Computational Science – ICCS 2025: 25th International Conference, Singapore, Singapore, July 7–9, 2025, Proceedings, Part IV},
pages = {327–335},
numpages = {9},
location = {Singapore, Singapore}
}

@inproceedings{brink2023thicket,
author = {Brink, Stephanie and McKinsey, Michael and Boehme, David and Scully-Allison, Connor and Lumsden, Ian and Hawkins, Daryl and Burgess, Treece and Lama, Vanessa and L\"{u}ttgau, Jakob and Isaacs, Katherine E. and Taufer, Michela and Pearce, Olga},
year = {2023},
month = {June},
title = {{Thicket: Seeing the Performance Experiment Forest for the Individual Run Trees}},
booktitle = {32nd Intl Symposium on High-Performance Parallel and Distributed Computing}
}

@inproceedings{beckingsaleRAJAPortablePerformance2019,
  title = {{{RAJA}}: {{Portable Performance}} for {{Large-Scale Scientific Applications}}},
  shorttitle = {{{RAJA}}},
  booktitle = {2019 {{IEEE}}/{{ACM International Workshop}} on {{Performance}}, {{Portability}} and {{Productivity}} in {{HPC}} ({{P3HPC}})},
  author = {Beckingsale, David A. and Burmark, Jason and Hornung, Rich and Jones, Holger and Killian, William and Kunen, Adam J. and Pearce, Olga and Robinson, Peter and Ryujin, Brian S. and Scogland, Thomas RW},
  year = {2019},
  month = nov,
  pages = {71--81},
  publisher = {{IEEE}},
  address = {{Denver, CO, USA}},
  doi = {10.1109/P3HPC49587.2019.00012},
  isbn = {978-1-72816-003-0},
  langid = {english}
}

@misc{llnlRAJAPerf,
  title = {{RAJA Performance Suite}},
  author = {{LLNL}},
  year = {2017},
  copyright = {BSD-3-Clause},
  howpublished = {\url{http://github.com/LLNL/RAJAPerf}}
}

@techreport{llnl_loops86,
  author        = "F. H. McMahon",
  title         = {{Livermore Fortran kernels: A computer test of numerical performance range}},
  institution   =  "UCRL-53724",
  place = {Lawrence Livermore National Laboratory, Livermore, CA, USA},
  year = {1986},
  month = {December}
}

@misc{polybench,
  title = {{{The Polyhedral Benchmark Suite}}},
  author = "Louis-Noel Pouchet",
  year = {2010},
  howpublished = {\url{https://web.cs.ucla.edu/~pouchet/software/polybench/}}
}

@TECHREPORT{McCalpin2007,
  author = {John D. McCalpin},
  title = {STREAM: Sustainable Memory Bandwidth in High Performance Computers},
  institution = {University of Virginia},
  year = {1991-2007},
  address = {Charlottesville, Virginia},
  note = {A continually updated technical report.
http://www.cs.virginia.edu/stream/},
  url = {http://www.cs.virginia.edu/stream/}
}

@article{LDMS,
author = {Agelastos, Anthony and Allan, Benjamin and Brandt, Jim and Cassella, Paul and Enos, Jeremy and Fullop, J. and Gentile, Ann and Monk, Steve and Naksinehaboon, Nichamon and Ogden, Jeff and Rajan, Mahesh and Showerman, Michael and Stevenson, Joel and Taerat, Narate and Tucker, Tom},
year = {2015},
month = {01},
pages = {154-165},
title = {The Lightweight Distributed Metric Service: A Scalable Infrastructure for Continuous Monitoring of Large Scale Computing Systems and Applications},
volume = {2015},
journal = {International Conference for High Performance Computing, Networking, Storage and Analysis, SC},
doi = {10.1109/SC.2014.18}
}

@inproceedings{10.1145/3736731.3746150,
author = {Pearce, Olga and Becker, Gregory and Brink, Stephanie and Hanford, Nathan and Yokelson, Dewi and Knox, August and Rountree, Barry},
title = {HPC Benchmarking: Repeat, Replicate, Reproduce},
year = {2025},
isbn = {9798400719585},
publisher = {Association for Computing Machinery},
address = {New York, NY, USA},
url = {https://doi.org/10.1145/3736731.3746150},
doi = {10.1145/3736731.3746150},
booktitle = {Proceedings of the 3rd ACM Conference on Reproducibility and Replicability},
pages = {85–95},
numpages = {11},
location = {
},
series = {ACM REP '25}
}

@article{papi, 
  author = {Browne, S. and others}, 
  title = {{A Portable Programming Interface for Performance Evaluation on Modern Processors}}, 
  year = {2000}, issue_date = {August 2000}, publisher = {Sage Publications, Inc.}, address = {USA}, 
  volume = {14}, number = {3}, issn = {1094-3420}, 
url = {https://doi.org/10.1177/109434200001400303}, 
doi = {10.1177/109434200001400303}, journal = {Int. J. High Perform. Comput. Appl.},
   month = {aug}, pages = {189-204}, numpages = {16} }

@inproceedings{boehme:2016:caliper,
author = {Boehme, David and Gamblin, Todd and Beckingsale, David and Bremer, Peer-Timo and Gimenez, Alfredo and LeGendre, Matthew and Pearce, Olga and Schulz, Martin},
title = {{Caliper: Performance Introspection for HPC Software Stacks}},
year = {2016},
isbn = {9781467388153},
publisher = {IEEE Press},
booktitle = {Proceedings of the International Conference for High Performance Computing, Networking, Storage and Analysis},
articleno = {47},
numpages = {11},
location = {Salt Lake City, Utah},
series = {SC '16}
}

@misc{caliper,
  title = {{{Caliper}}},
  author = {{LLNL}},
  year = {2017},
  copyright = {BSD 3-clause},
  howpublished = {\url{https://github.com/llnl/caliper}}
}

@misc{thicket,
  title = {{{Thicket}}},
  author = {{LLNL}},
  year = {2023},
  copyright = {MIT},
  howpublished = {\url{https://github.com/llnl/thicket}}
}

@inproceedings{yasin2014top,
  title = {{A {{Top-Down}} Method for Performance Analysis and Counters Architecture}},
  booktitle = {2014 {{IEEE International Symposium}} on {{Performance Analysis}} of {{Systems}} and {{Software}} ({{ISPASS}})},
  author = {Yasin, Ahmad},
  year = {2014},
  month = mar,
  pages = {35--44},
  publisher = {{IEEE}},
  address = {{CA, USA}},
  doi = {10.1109/ISPASS.2014.6844459},
  isbn = {978-1-4799-3606-9 978-1-4799-3604-5},
  langid = {english}
}

@misc{ncu,
Howpublished = {\url{https://docs.nvidia.com/nsight-compute/NsightCompute/index.html }},
Title = {{NVIDIA Nsight Compute Profiling Tool}},
author = {NVIDIA},
}

@INPROCEEDINGS{gpu_rooflines,
  author={Ding, Nan and Williams, Samuel},
  booktitle={2019 IEEE/ACM Performance Modeling, Benchmarking and Simulation of High Performance Computer Systems (PMBS)}, 
  title={An Instruction Roofline Model for GPUs}, 
  year={2019},
  volume={},
  number={},
  pages={7-18},
  doi={10.1109/PMBS49563.2019.00007}
}

@misc{scikitlearnAgglom,
  title = {{{AgglomerativeClustering}}},
  year = {2024},
  author = {scikit learn},
  howpublished = {https://scikit-learn.org/stable/modules/generated/sklearn.cluster.AgglomerativeClustering.html}
}

@misc{scikitlearnKmeans,
    title = {{Kmeans}},
    year = {2026},
    author = {scikit learn},
    howpublished = {https://scikit-learn.org/stable/modules/generated/sklearn.cluster.KMeans.html}
}

@inproceedings{10.5555/1283383.1283494,
author = {Arthur, David and Vassilvitskii, Sergei},
title = {k-means++: the advantages of careful seeding},
year = {2007},
isbn = {9780898716245},
publisher = {Society for Industrial and Applied Mathematics},
address = {USA},
booktitle = {Proceedings of the Eighteenth Annual ACM-SIAM Symposium on Discrete Algorithms},
pages = {1027–1035},
numpages = {9},
location = {New Orleans, Louisiana},
series = {SODA '07}
}

@article{ward_strat_agglom,
 URL = {http://www.jstor.org/stable/2282967},
 author = {Joe H. Ward},
 journal = {Journal of the American Statistical Association},
 number = {301},
 pages = {236--244},
 publisher = {[American Statistical Association, Taylor & Francis, Ltd.]},
 title = {Hierarchical Grouping to Optimize an Objective Function},
 volume = {58},
 year = {1963}
}

@article{richards:ecp2021,
  author       = {D. F. Richards and O. Aaziz and 
                  J. Cook and J. Kuehn and G. Watson and
                  P. McCorquodale and W. Godoy and
                  J. Delozier and M. Carroll and
                  C. Vaughan
},
  title        = {Quantitative Performance Assessment of Proxy Apps and Parents Report for ECP Proxy App Project Milestone ADCD-504-11},
  year         = {2021},
  url          = {https://www.osti.gov/servlets/purl/1860797},
}

@article{f0f3240e4a9510cc07515899834b8e7b122f1a38,
title = {Quantifying Locality In The Memory Access Patterns of HPC Applications},
year = {2005},
url = {https://www.semanticscholar.org/paper/f0f3240e4a9510cc07515899834b8e7b122f1a38},
author = {Jonathan Weinberg and M. O. McCracken and Erich Strohmaier and A. Snavely},
journal = {ACM/IEEE SC 2005 Conference (SC'05)},
volume = {null},
pages = {50-50},
doi = {10.1109/SC.2005.59},
}

@inproceedings{matching-signatures, author = {Jayakumar, Anirudh and Murali, Prakash and Vadhiyar, Sathish}, title = {Matching Application Signatures for Performance Predictions Using a Single Execution}, year = {2015}, isbn = {9781479986491}, publisher = {IEEE Computer Society}, address = {USA}, url = {https://doi.org/10.1109/IPDPS.2015.20}, doi = {10.1109/IPDPS.2015.20}, booktitle = {Proceedings of the 2015 IEEE International Parallel and Distributed Processing Symposium}, pages = {1161–1170}, numpages = {10}, series = {IPDPS '15} }


\end{document}